\documentclass[10pt,journal,compsoc]{IEEEtran}

\ifCLASSOPTIONcompsoc
  \usepackage[nocompress]{cite}
\else
  \usepackage{cite}
\fi

\ifCLASSINFOpdf
  \usepackage[pdftex]{graphicx}
  \graphicspath{./figures/}
  \DeclareGraphicsExtensions{.pdf,.jpeg,.png}
\else
  \usepackage[dvips]{graphicx}
  \graphicspath{{../eps/}}
  \DeclareGraphicsExtensions{.eps}
\fi
\usepackage{amsmath}
\usepackage{amssymb}
\usepackage{mathtools}
\usepackage{url}

\usepackage{breakurl}
\usepackage[breaklinks]{hyperref}

\usepackage{algorithm}
\usepackage[noend]{algpseudocode}

\DeclarePairedDelimiter{\floor}{\lfloor}{\rfloor}

\begin{document}
\title{SPRING: A Sparsity-Aware Reduced-Precision Monolithic 3D CNN Accelerator \\ Architecture for 
Training and Inference}

\author{Ye~Yu,
        and~Niraj~K.~Jha,~\IEEEmembership{Fellow,~IEEE}
\IEEEcompsocitemizethanks{\IEEEcompsocthanksitem This work was supported by NSF under Grant No.
CCF-1811109. Y. Yu and N.K. Jha are with the Department
of Electrical Engineering, Princeton University, Princeton,
NJ, 08540.\protect\\
E-mail: \{yeyu,jha\}@princeton.edu}%
}

\markboth{}%
{Yu \MakeLowercase{\textit{et al.}}: SPRING: A Sparsity-Aware Reduced-Precision Monolithic 3D CNN 
Accelerator Architecture for Training and Inference}

\IEEEtitleabstractindextext{%
\begin{abstract}
Convolutional neural networks (CNNs) outperform traditional machine learning algorithms across a 
wide range of applications, such as object recognition, image segmentation, and autonomous driving. 
However, their ever-growing computational complexity makes it necessary to design efficient 
hardware accelerators.  Most CNN accelerators focus on exploring various dataflow styles 
and designs that exploit computational parallelism. However, potential performance improvement from 
sparsity (in activations and weights) has not been adequately addressed. The computation and 
memory footprint of CNNs can be 
significantly reduced if sparsity is exploited in network evaluations. Therefore, different 
pruning methods have been proposed to increase sparsity. To take advantage of sparsity, some 
accelerator designs explore sparsity encoding and evaluation on CNN accelerators. However, sparsity 
encoding is just performed on activation data or CNN weights and only used in inference. It has been 
shown that activations and weights also have high sparsity levels during the network training phase. 
Hence, sparsity-aware computation should also be considered in the training phase.  To further 
improve performance and energy efficiency, some accelerators evaluate CNNs with limited precision. 
However, this is limited to the inference phase since reduced precision sacrifices network accuracy 
if used in training.  In addition, CNN evaluation is usually memory-intensive, especially 
during training. The performance bottleneck arises from the fact that the memory cannot 
feed the computational units enough data, resulting in idling of these computational units 
and thus low utilization ratios. A 3D memory interface has been used on high-end GPUs to alleviate 
memory bandwidth shortage.  In this article, we propose SPRING, a SParsity-aware Reduced-precision 
Monolithic 3D CNN accelerator for trainING and inference. SPRING supports both CNN training and 
inference.  It uses a binary mask scheme to encode sparsities in activations and weights. It 
uses the stochastic rounding algorithm to train CNNs with reduced precision without accuracy loss.
To alleviate the memory bottleneck in CNN evaluation, especially during training, SPRING uses an
efficient monolithic 3D nonvolatile memory interface to increase memory bandwidth. Compared to
Nvidia GeForce GTX 1080 Ti, SPRING achieves 15.6$\times$, 4.2$\times$, and 66.0$\times$ 
improvements in performance, power reduction, and energy efficiency, respectively, for CNN training, 
and 15.5$\times$, 4.5$\times$, and 69.1$\times$ improvements in performance, power reduction, and 
energy efficiency, respectively, for inference.
\end{abstract}

}

\maketitle

\IEEEdisplaynontitleabstractindextext
\IEEEpeerreviewmaketitle

\ifCLASSOPTIONcompsoc
\IEEEraisesectionheading{\section{Introduction}\label{introduction}}
\else
\section{Introduction}
\label{introduction}
\fi

\IEEEPARstart{C}{onvolutional} neural networks (CNNs) excel at various important applications,
e.g., image classification, image segmentation, robotics control, and natural language processing. 
However, their high computational complexity necessitates specially-designed accelerators for 
efficient processing. Training of CNNs requires an enormous amount of computing power to 
automatically learn the weights based on a large training dataset. Few ASIC-based CNN training 
accelerators have been presented \cite{dadiannao,scaledeep,tpuv2}.  However, graphical
processing units (GPUs) typically play a dominant in the training phase as CNN computation 
essentially maps well to their single-instruction multiple-data (SIMD) units and the large number 
of SIMD units present in GPUs provide significant computational throughput for training CNNs 
\cite{pascal,volta}. In addition, the higher clock speed, bandwidth, and power management 
capabilities of the Graphics Double Data Rate (GDDR) memory relative to the regular DDR memory 
make GPUs the de facto accelerator choice for CNN training. On the other hand, CNN inference is more 
latency- and power-sensitive as an increasing number of applications need real-time CNN evaluations 
on battery-constrained edge devices. Hence, ASIC- and FPGA-based accelerators have been widely 
explored for this purpose \cite{centric,escher,scalpel,bit,maeri}. However, they can only process low-level CNN operations, such as convolution and matrix multiplication, and lack the 
flexibility of a general-purpose processor. Although CNN models have evolved rapidly recently, their 
fundamental building blocks are common and long-lasting. Therefore, the ASIC- and FPGA-based 
accelerators can efficiently process new CNN models with their domain-specific architectures. 
FPGA-based accelerators achieve faster time-to-market and enable prototyping of new accelerator 
designs. Microsoft has used customized FPGA boards, called Catapult \cite{catapult}, in its data 
centers to accelerate Bing ranking by 2$\times$. An FPGA-based CNN accelerator that uses on-chip 
memory has been proposed in \cite{only}, where a fixed-point representation is used to keep all the 
weights stored in on-chip memory thus avoiding the need to access external memory. To improve 
dynamic resource utilization of FPGA-based CNN accelerators, multiple accelerators, each specialized 
for a specific CNN layer, have been constructed using the same FPGA hardware resource 
\cite{underutilization}. A convolver design for both the convolutional (CONV) layer and 
fully-connected (FC) layer has been proposed in \cite{embedded} to efficiently process CNNs on 
embedded FPGA platforms. ASIC-based CNN accelerators have better energy efficiency and can be fully 
customized for CNN applications. In \cite{shidiannao}, CNNs are mapped entirely within the on-chip 
memory and the ASIC accelerator is placed close to the image sensor so that all the DRAM accesses are 
eliminated, leading to a 60$\times$ energy efficiency improvement relative to previous works. A 
1D chain ASIC architecture is used in \cite{chain} to accelerate the CONV layers since these
layers are the most compute-intensive \cite{dadiannao}. To speed up the CONV layers, a Fast Fourier 
Transform-based fast multiplication is used in \cite{circnn}. This accelerator encodes weights 
using block-circulant matrices and converts convolutions into matrix multiplications to reduce 
the computational complexity from $O(n^2)$ to $O(nlogn)$ and storage complexity from $O(n^2)$ to 
$O(n)$. A 3D memory system is used in \cite{tetris} to reduce memory bandwidth pressure. This
enables more chip area to be devoted to processing elements (PEs), thus increasing performance and 
energy efficiency.

To take advantage of the underlying parallel computing resources of CNN accelerators, an efficient 
dataflow is necessary to minimize data movement between the on-chip memory and PEs. Unlike the
temporal architectures, like SIMD or single-instruction multiple-thread, used in 
central processing units (CPUs) and GPUs, the Google Tensorflow processing unit (TPU) uses a 
spatial architecture, called the systolic array \cite{tpu}. Data flow into arithmetic logic units 
(ALUs) in a wave and move through adjacent ALUs in order to be reused. In \cite{fuse}, multiple 
CNN layers are fused and processed so that intermediate data can be kept in on-chip memory,
thus obviating the need for external memory access. A fine-grained dataflow accelerator is proposed 
in \cite{fine}. It converts convolution into data preprocessing and matrix multiplication. Data are 
directly transferred among PEs, without the need for redundant control logic, as opposed to temporal 
architectures, such as DianNao \cite{diannao}. In \cite{flex}, a flexible dataflow architecture is 
described for efficiently processing different types of parallelism: feature map, neuron, and 
synapse. A dataflow called row-stationary is used in \cite{eyeriss} to reuse data and minimize data 
movement on a spatial architecture.

Although various dataflow styles and computational parallelism designs have been explored in recent 
works, the potential speedup from weight/activation sparsity is still underexplored. The computation and 
memory footprint of CNNs can be significantly reduced if sparsity is exploited during network 
evaluations. Some recent works utilize sparsity to speed up CNN evaluations 
\cite{eie,sparsenn,scnn,ucnn,cnvlutin,cambriconx}. However, they only consider either activation 
or weight sparsity, and only use sparsity during CNN inference based on various pruning methods. 
It has been shown that the average network-wide activation sparsity of the well-known AlexNet 
CNN \cite{alexnet} during its entire training process is 62\% (a maximum of 93\%) \cite{dma}. 
Therefore, the training process can be significantly accelerated if sparsity is exploited. 
Another CNN acceleration technique is to use reduced precision to improve performance and energy 
efficiency. For example, TPU and DianNao use 8-bit and 16-bit fixed-point quantizations, 
respectively, in CNN evaluations. However, low-precision accelerators are currently mainly used in 
the inference phase, since CNN training involves gradient computation and propagation that require 
high-precision floating-point operations to achieve high accuracy. Apart from improving the 
efficiency of the computational resources employed in CNN accelerators, CNN training also requires 
a large memory bandwidth to store activations and weights. In the forward pass, activations must be 
retained in the memory until the backward pass commences in order to compute the error gradients 
and update weights. Besides, in order to fill the SIMD units of GPUs, a large amount of data is 
needed from the memory. Hence, 3D memory systems, such as hybrid memory cube (HMC) \cite{HMC} and 
high bandwidth memory (HBM) \cite{HBM}, have been used in high-end GPUs
to provide significant
memory bandwidth for CNN training and to bring processing closer to computing. 
There are many studies on accelerators with near-memory processing (NMP) or 
processing in memory (PIM) that aim to reduce the memory transfer overhead 
\cite{near,chameleon,prime,neurocube,isaac,time,nnpim}. In \cite{tensordimm}, 
an NMP-enhanced dual in-line memory module is proposed to reduce the latency 
of embedding fetching and gather/reduction operations used in recommender 
deep neural networks (DNNs) that take up 34\% of the total execution time of 
DNN workloads in Facebook datacenters \cite{fb}. A PIM accelerator, FloatPIM, 
is proposed in \cite{floatpim}. It speeds up data movement by enabling 
parallel data transfer between neighboring blocks.

In this article, we make the following contributions:
\newline
\indent1) We propose a novel sparsity-aware CNN accelerator architecture, called SPRING. It encodes 
activation and weight sparsities with binary masks and uses efficient low-overhead hardware 
implementations for CNN training and inference.
\newline
\indent2) SPRING uses reduced-precision fixed-point operations for both training and inference. A 
dedicated module is used to implement the stochastic rounding algorithm \cite{sr} to prevent 
accuracy loss during CNN training. 
\newline
\indent3) SPRING uses an efficient monolithic 3D nonvolatile RAM (NVRAM) interface to provide 
significant memory bandwidth for CNN processing. This alleviates the performance bottleneck in 
CNN training since the training process is usually memory-bound \cite{bound}.

We test the proposed SPRING architecture on seven well-known CNNs in the context of both training 
and inference. Simulation results show that the average execution time, power dissipation, and 
energy consumption are reduced by 15.6$\times$, 4.2$\times$, and 66.0$\times$, respectively, for 
CNN training, and 15.5$\times$, 4.5$\times$, and 69.1$\times$, respectively, for inference, relative 
to Nvidia GeForce GTX 1080 Ti. 

The rest of the article is organized as follows. Section~\ref{background} discusses the
background information required to understand our sparsity-aware accelerator. 
Section~\ref{architecture} presents the sparsity-aware reduced-precision accelerator architecture.
Section~\ref{simulations} describes our simulation setup and flow. Section~\ref{results}
presents experimental results obtained on seven typical CNNs. Section~\ref{discussion}
discusses the limitations of our work. Section~\ref{conclusion} concludes the article.

\section{Background}
\label{background}
In this section, we discuss the background material necessary for understanding our proposed 
sparsity-aware reduced-precision accelerator architecture. We first give a primer on CNNs. We then 
discuss existing sparsity-aware designs. Then, we discuss various CNN training algorithms 
that use low numerical precision. Finally, we describe an efficient on-chip memory interface that 
is used for CNN acceleration.

\subsection{CNN overview}
\begin{figure}[!t]
\centering
\includegraphics[width=3.5in]{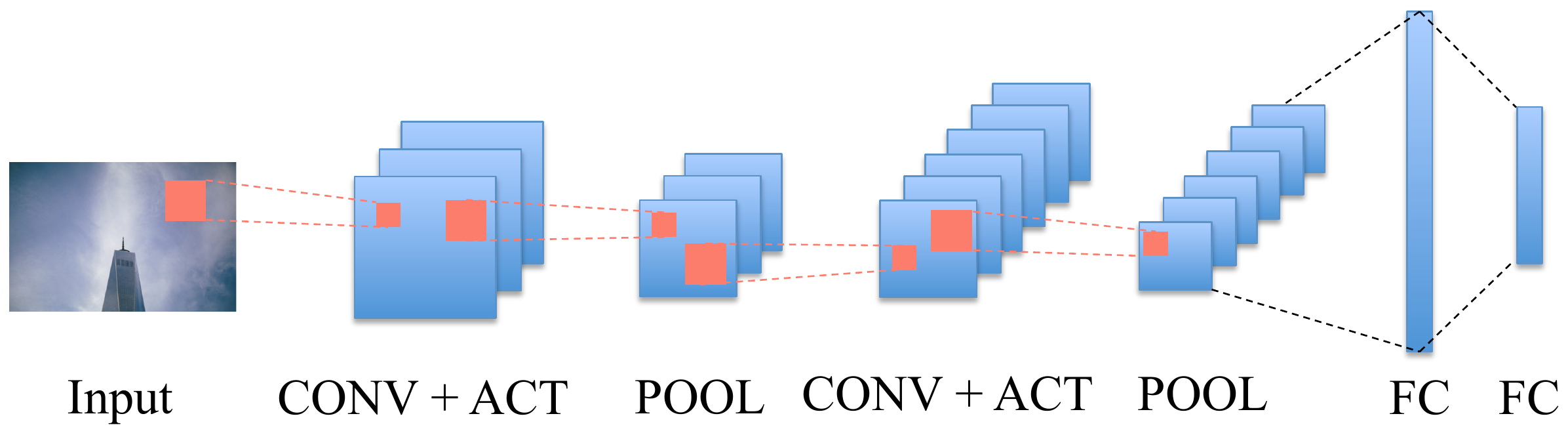}
\caption{CNN architecture illustration}
\label{CNN}
\end{figure}

Although different CNNs have different hyperparameters, such as the number of layers and shapes, 
they share a similar architecture, as shown in Fig.~\ref{CNN}. CNNs are generally composed of five 
building blocks: CONV layers, activation (ACT) layers, pooling (POOL) layers, batch normalization 
layers (not shown in Fig.~\ref{CNN}), and FC layers. Among these basic components, the CONV and FC 
layers are the most compute-intensive \cite{dadiannao}.  We describe them next.

\textbf{CONV layers:} A batch of 3D input feature maps is convolved with a 
set of 3D filter weights 
to generate a batch of 3D output feature maps. The filter weights are usually fetched from 
external memory once and stored in on-chip memory as they are shared among multiple convolution 
windows. Therefore, CONV layers have relatively low memory bandwidth pressure and are usually 
compute-bound as they require a large number of convolution computations. Given the input feature 
map \textbf{I} and filter weights \textbf{W}, the output feature map \textbf{O} is computed as follows:

\begin{align}
\begin{split}\label{eq:1}
    \textbf{O}[n][k][p][q] = {}&\sum\limits_{c=0}^{C-1}\sum\limits_{r=0}^{R-1}\sum\limits_{s=0}^{S-1}\textbf{I}[n][c][p\times u+r][q\times v+s]\\
    & \times \textbf{W}[k][c][r][s]
\end{split}
\end{align}

\noindent where \textbf{I} $\in \mathbb{R}^{NCHW}$, \textbf{W} $\in \mathbb{R}^{KCRS}$, and 
\textbf{O} $\in \mathbb{R}^{NKPQ}$. $N$ is the number of images in a batch and $K$ is the total 
number of filters in the CONV layer. $C$ represents the number of channels in the input feature maps 
and filter weights. $H$ and $W$ denote the height and width of the input feature maps, respectively, 
whereas $R$ and $S$ denote the height and width of filter weights, respectively. The vertical 
and horizontal strides are given by $u$ and $v$, respectively. The height and width of the output 
feature maps are given by $P$ and $Q$, respectively.

\textbf{FC layers:} The neurons in an FC layer are fully connected with neurons in the previous layer 
with a specific weight associated with each connection. It is the most memory-intensive layer in 
CNNs \cite{embedded,caffeine} since no weight is reused. The computation of the FC layer can be 
represented by a matrix-vector multiplication as follows:  

\begin{equation}
    \textbf{y}=\textbf{Wx}+\textbf{b}
\end{equation}

\noindent where \textbf{W} $\in \mathbb{R}^{m\times n}$, \textbf{y}, \textbf{b} $\in \mathbb{R}^m$, 
and \textbf{x} $\in \mathbb{R}^n$. The output and input neurons of the FC layer are represented in 
vector form as $\textbf{y}$ and $\textbf{x}$. $\textbf{W}$ represents the weight matrix and 
$\textbf{b}$ is the bias vector associated with the output neurons.

\subsection{Exploiting sparsity in CNN accelerators}
It is known that the sparsity levels of CNN weights typically range from 20\% to 80\%
\cite{deepcompression,learning}, and when the rectified linear unit (ReLU) activation function
is employed, the activations are clamped to zeros in the 50\% to 70\% 
range \cite{scnn}. The 
combination of weight and activation sparsities can reduce computations and memory accesses 
significantly if the accelerator can support sparsity-aware operations. In order to speed up
CNN evaluation by utilizing weight/activation sparsity, the first step is to encode the sparse data 
in a compressed format that can be efficiently processed by accelerators.  EIE is an
accelerator that encodes a sparse weight matrix in a compressed sparse column (CSC) format 
\cite{csc} and uses a second vector to encode the number of zeros between adjacent non-zero 
elements \cite{eie}. However, it is only used to speed up the FC layers and has no impact on 
the CONV layers. Hence, a majority of CNN computations does not benefit from sparsity-aware 
acceleration.  A lightweight run-time output sparsity predictor has been developed in SparseNN, an 
architecture enhanced from EIE, to accelerate CNN inference \cite{sparsenn}. Activations in 
the CSC format are first fed to the lightweight predictor to predict the non-zero elements in the 
output neurons. Then, the activations associated with non-zero outputs are sent to feedforward 
computations to bypass computations that lead to zero outputs. If the number of computations skipped 
is large enough, the overhead of output predictions can be offset. However, since the output sparsity 
predicted by the lightweight predictor is an approximation of the real sparsity value, it 
incurs an accuracy loss that makes it unsuitable for CNN training.  SCNN is another accelerator
that uses a zero-step format to encode weight/activation sparsity: an index vector is used to indicate 
the number of non-zero data points and the number of zeros before each non-zero data point. It 
multiplies activation and weight vectors in a manner similar to a Cartesian product using an input 
stationary dataflow \cite{scnn}. However, the Cartesian product does not automatically align non-zero 
weights and activations in the FC layers since the FC layer weights are not reused as in 
the case of CONV layers. This leads to performance degradation for FC layers and makes SCNN 
unattractive for CNNs dominated by FC layers. Stitch-X \cite{stitch}, an improved version of SCNN, adopts a hybrid dataflow by leveraging both spatial and temporal partial-sum reduction to dynamically stitch together non-zero activations and weights. Cnvlutin \cite{cnvlutin} enhances the DaDianNao 
architecture to support zero-skipping in activations using a zero-step offset vector that is similar 
to graphics processor proposals \cite{warp,simt,level,convergence}. The limitation of this 
architecture is that the length of offset vectors in different PEs may be different.  Hence, they 
may require different numbers of cycles to process the data. Thus, the PE with the longest offset vector 
becomes the performance bottleneck while other PEs idle and wait for it.  Cambricon-X is an
accelerator that also employs the zero-step sparsity encoding method and uses a dedicated indexing 
module to select and transfer needed neurons to PEs, with a reduced memory bandwidth requirement 
\cite{cambriconx}. The PEs run asynchronously to avoid the idling problem of Cnvlutin. An enhanced version, Cambricon-S, is then proposed to reduce the irregularity of weight sparsity using a software-based
coarse-grained pruning technique \cite{cambricons}.
UCNN is an accelerator that improves CNN inference performance by exploiting weight repetition in 
the CONV layers \cite{ucnn}. It uses a factorized dot product dataflow to reduce the number of 
multiplications and a memorization method to reduce weight memory access via weight repetition.

Both the CSC and zero-step encoding formats compress data by eliminating zero-elements and the 
accelerators discussed above efficiently process the compressed data. However, weight/activation 
sparsity can not only be exploited at the PE level but also at the bit level. Stripes, a 
bit-serial hardware accelerator, avoids the processing of zero prefix and suffix bits through 
serial-parallel multiplications on CNNs \cite{stripes}. Each bit of a neuron is processed at every 
cycle and zero bits are skipped on the fly. Multiple neurons are processed in parallel to mitigate 
performance loss from bit-serial processing. Pragmatic, a CNN accelerator enhanced from Stripes, 
supports zero-bit skipping regardless of its position \cite{pragmatic}. However, it needs to 
convert the input neuron representation into a zero-bit-only format on the fly, which leads to up to 
a 16-cycle latency.

\subsection{Low-precision CNN training algorithms}
The rapid evolution of CNNs in recent years has necessitated the deployment of large-scale 
distributed training using high-performance computing infrastructure \cite{gpu,distributed,cluster}. 
Even with such a powerful computing infrastructure, training a CNN to convergence usually takes 
several days, sometimes even a few weeks. Hence, to speed up the CNN training process, various 
training algorithms with low-precision computations have been proposed.

Single-precision floating-point (FP32) operation has mainly been used as the training standard on 
GPUs. Meanwhile, efforts have been made to train CNNs with half-precision floating-point (FP16) 
arithmetic since it can improve training throughput by 2$\times$, in theory, on the same computing 
infrastructure. However, compared to FP32, FP16 involves rounding off gradient values and quantizing 
to a lower-precision representation. This introduces noise in the training process and defers CNN 
convergence. To maintain a balance between the convergence rate and training throughput, 
mixed-precision training algorithms that use a combination of FP32 and FP16 have been proposed 
\cite{mixed,micikevicius2018mixed}. The FP16 representation is used in the most compute-intensive 
multiplications and the results are accumulated into FP32 outputs. Dynamic scaling is required to 
prevent the vanishing gradient problem \cite{VGP}.

Compared to floating-point arithmetic, fixed-point operations are much faster 
and more energy-efficient 
on hardware accelerators, but have a lower dynamic range. To overcome the dynamic range limitation, 
the dynamic fixed-point format \cite{dynamic} is used in CNN training \cite{training,das2018mixed}. 
Unlike the regular fixed-point format, the dynamic fixed-point format uses multiple scaling factors 
that are updated during training to adjust the dynamic range of different groups of variables. The 
CNN training convergence rate is highly sensitive to the rounding scheme used in fixed-point 
arithmetic \cite{sr}. Instead of tuning the dynamic range used in the dynamic fixed-point format, 
a stochastic rounding method has been proposed to leverage the noise tolerance of CNN algorithms 
\cite{sr}. CNNs are trained in a manner that the rounding error is exposed to the network and weights 
are updated accordingly to mitigate this error, without impacting the convergence rate.

\subsection{Efficient on-chip memory interface and emerging NVRAM technologies}
CNN training involves feeding vast input feature maps and filter weights to the accelerator computing 
units to compute the error gradients used to update CNN weights in backpropagation. Besides the 
large memory size required to store all the CNN weights, a high memory bandwidth becomes 
indispensable to keep running the computing units at full throughput. Hence, through-silicon via 
(TSV)-based 3D memory interfaces have been used on high-end GPUs \cite{volta} and specialized CNN 
accelerators \cite{tpuv2}. The most widely-used TSV-based 3D memory interface is HBM. In each HBM 
package, multiple DRAM dies and one memory controller die are first fabricated and tested 
individually. Then, these dies are aligned, thinned, and bonded using TSVs. The HBM package is 
connected to the processor using an interposer in a 2.5D manner. This shortens the interconnects 
within the memory system and between the memory and processor, thus reducing memory access latency 
and improving memory bandwidth. In addition, since more DRAM dies are integrated within the same 
footprint area, HBM enables smaller form factors: HBM-2 uses 94\% less space relative to GDDR5 for 
a 1GB memory \cite{hbm2}.

Apart from improving the DRAM interface, the industry has also been exploring various NVRAM 
technologies to replace DRAM, such as ferroelectric RAM (FeRAM), spin-transfer torque magnetic RAM 
(STT-MRAM), phase-change memory (PCM), nanotube RAM (NRAM), and resistive RAM (RRAM). It has
been shown in \cite{monolithic,hybrid} that RRAM can be used in an efficient 3D memory interface to 
deliver high memory bandwidth and energy efficiency. Information is represented by different 
resistance levels in an RRAM cell. Compared to a DRAM, an RRAM cell needs a higher current to change 
its resistance level. Therefore, the access transistors of an RRAM are larger than those of 
a DRAM \cite{5ns}. However, DRAM is expected to reach the scaling limit at 16$nm$ \cite{16nm} 
whereas RRAM is believed to be suitable for sub-10$nm$ nodes \cite{10nm}. Hence, the smaller 
technology node of an RRAM should offset the access transistor overhead. Besides, the nonvolatility of 
RRAM eliminates the need for a periodic refresh that a DRAM requires. This not only saves energy and 
reduces latency, but also gets rid of the refresh circuitry used in DRAM.

\section{Sparsity-aware reduced-precision~accelerator architecture}
\label{architecture}
In this section, we present the proposed architecture, SPRING: a sparsity-aware reduced-precision 
CNN accelerator for both training and inference. We first discuss accelerator architecture design 
and then dive into sparsity-aware acceleration, reduced-precision processing, and the monolithic 
3D NVRAM interface. 

\begin{figure}[!t]
\centering
\includegraphics[width=3.5in]{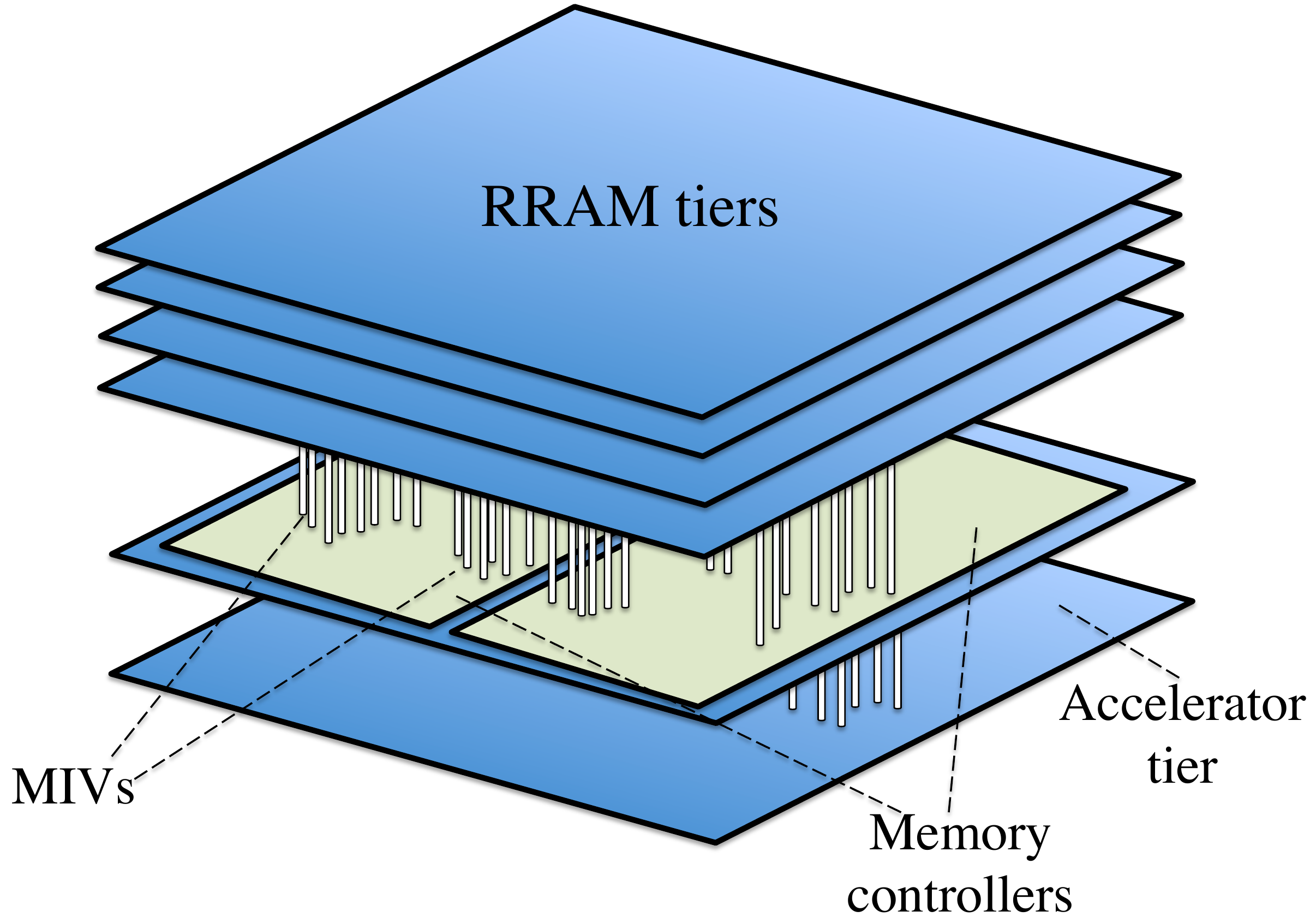}
\caption{The SPRING architecture}
\label{spring}
\end{figure}

Fig.~\ref{spring} shows the high-level view of the architecture. SPRING uses monolithic 3D 
integration to connect the accelerator tier with an RRAM interface. Unlike TSV-based 3D integration, 
monolithic 3D integration only has one substrate wafer, where devices are fabricated tier over tier. 
Hence, the alignment, thinning, and bonding steps of TSV-based 3D integration can be eliminated. In 
addition, tiers are connected through monolithic inter-tier vias (MIVs), whose diameter is the same 
as that of local vias and one-to-two orders of magnitude smaller than that of TSVs. This enables
a much higher MIV density ($10^8/mm^2$ at 14$nm$ \cite{density}), thus leaving much more space for 
logic. The accelerator tier is put at the bottom, on top of which is the memory controller tier. 
Above the memory controller tier lie the multiple RRAM tiers.

\begin{figure}[!t]
\centering
\includegraphics[width=3.5in]{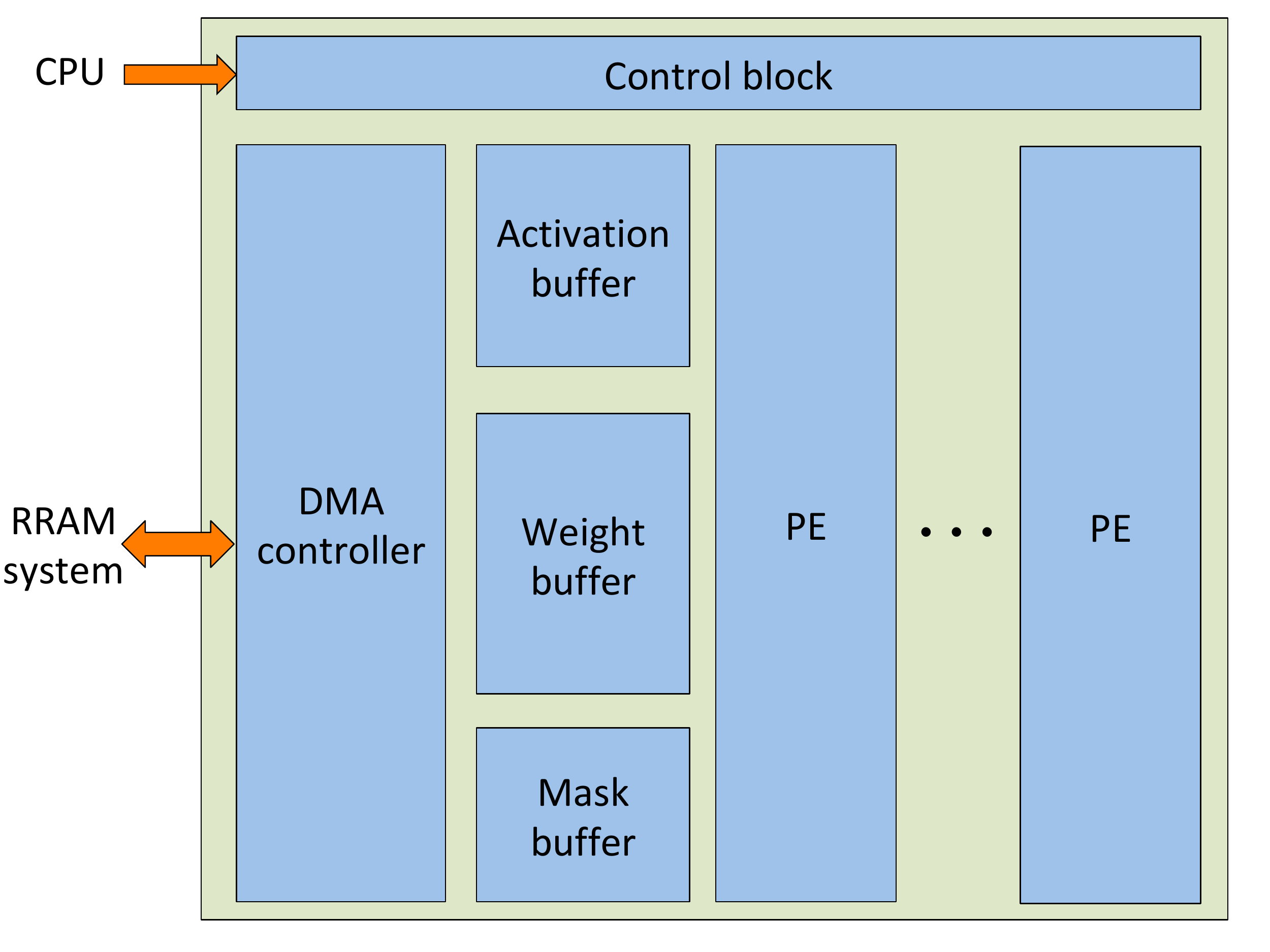}
\caption{Accelerator organization}
\label{organization}
\end{figure}

Fig.~\ref{organization} shows the organization of the accelerator tier. The  control block handles 
the CNN configuration sent from the CPU. It fetches the instruction stream and controls the rest of 
the accelerator to perform acceleration. The activations and filter weights are brought on-chip 
from the RRAM system by a direct memory access (DMA) controller. Activations and weights are stored 
in the activation buffer and weight buffer, respectively, in a compressed format. Data compression 
relies on binary masks that are stored in a dedicated mask buffer. The compression scheme is 
discussed in Section~\ref{sparsity}. The compressed data and the associated masks are used in
the PEs for CNN evaluation. The PEs are designed to operate in parallel to maximize overall throughput.

\begin{figure}[!t]
\centering
\includegraphics[width=3.5in]{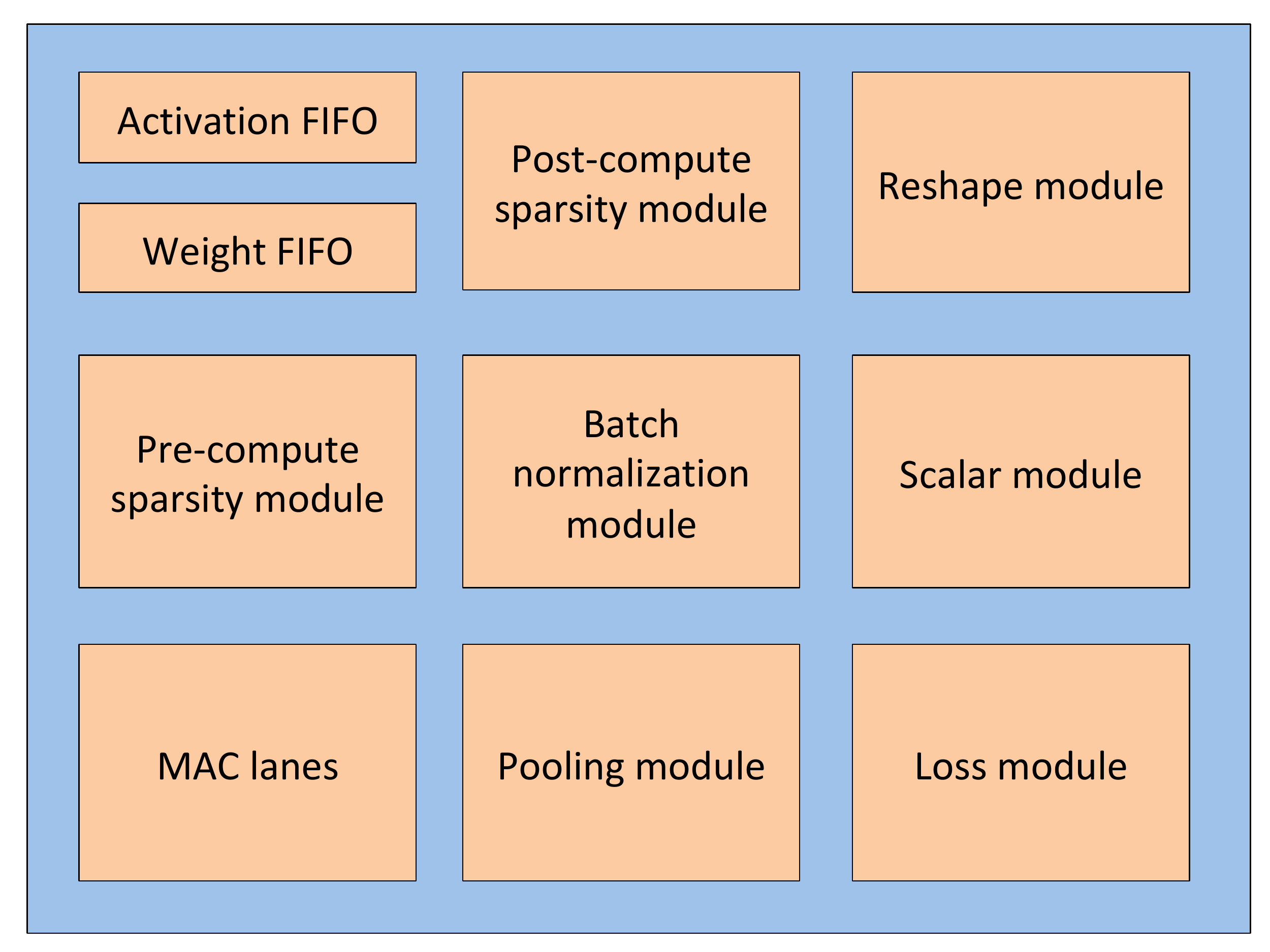}
\caption{Internal components of a PE}
\label{PE}
\end{figure}

Fig.~\ref{PE} shows the main components of a PE. The compressed data are buffered by the activation 
FIFO and weight FIFO. Then, they enter the pre-compute sparsity module along with the binary masks. 
Multiple multiplier-accumulator (MAC) lanes are used to compute convolutions or matrix-vector 
multiplications using zero-free activations and weights after they are preprocessed by the 
pre-compute sparsity module. The output results go through a post-compute sparsity module to maintain 
the zero-free format. Batch normalization operations \cite{batch} are used in modern CNNs to reduce 
the covariance shift. They are executed in the batch normalization module that supports both the 
forward pass and backward pass of batch normalization. Three pooling methods are supported by the 
pooling module: max pooling, min pooling, and mean pooling. The reshape module deals with matrix 
transpose and data reshaping. Element-wise arithmetic, such as element-wise add and subtract, is 
handled by the scalar module. Lastly, a dedicated loss module is used to process various loss 
functions, such as L1 loss, L2 loss, softmax, etc.

\subsection{Sparsity-aware acceleration}
\label{sparsity}
Traditional accelerator designs can only process dense data and do not support sparse-encoded 
computation. They treat zero elements in the same manner as regular data and thus perform operations 
that have no impact on the CNN evaluation results. In this context, weight/activation sparsity cannot 
be used to speed up computation and reduce the memory footprint.  In order to utilize sparsity to 
skip ineffectual activations and weights, and reduce the memory footprint, SPRING uses a binary-mask 
scheme to encode the sparse data and performs computations directly in the encoded format.

\begin{figure}[!t]
\centering
\includegraphics[width=3.5in]{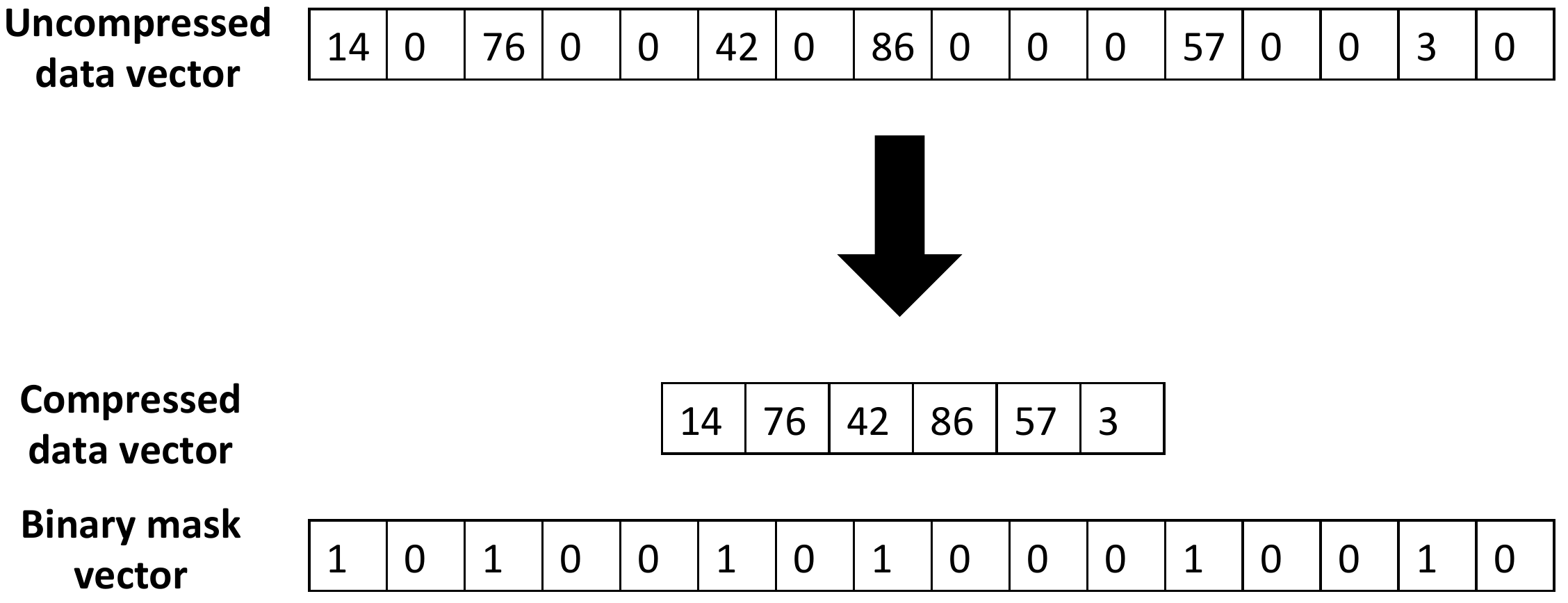}
\caption{The binary mask scheme: An example}
\label{binary}
\end{figure}

Compared to the regular dense format, SPRING compresses data vectors by removing all the
zero-elements. In order to retain the shape of the uncompressed data, an extra binary mask is
used. The binary mask has the same shape as that of the uncompressed data where each binary bit
in the mask is associated with one element in the original data vector. Fig.~\ref{binary} shows
an example of the binary-mask scheme that SPRING uses to compress activations and weights. The 
original uncompressed data vector has 16 elements, and if each element is represented using 16 bits, 
the total data length is 256 bits. With the binary scheme, only the six non-zero elements remain. The 
total length of the compressed data vector and the binary mask is 112 bits, which leads to a 
compression ratio of 2.3$\times$ for this example.

\begin{figure}[!t]
\centering
\includegraphics[width=3.5in]{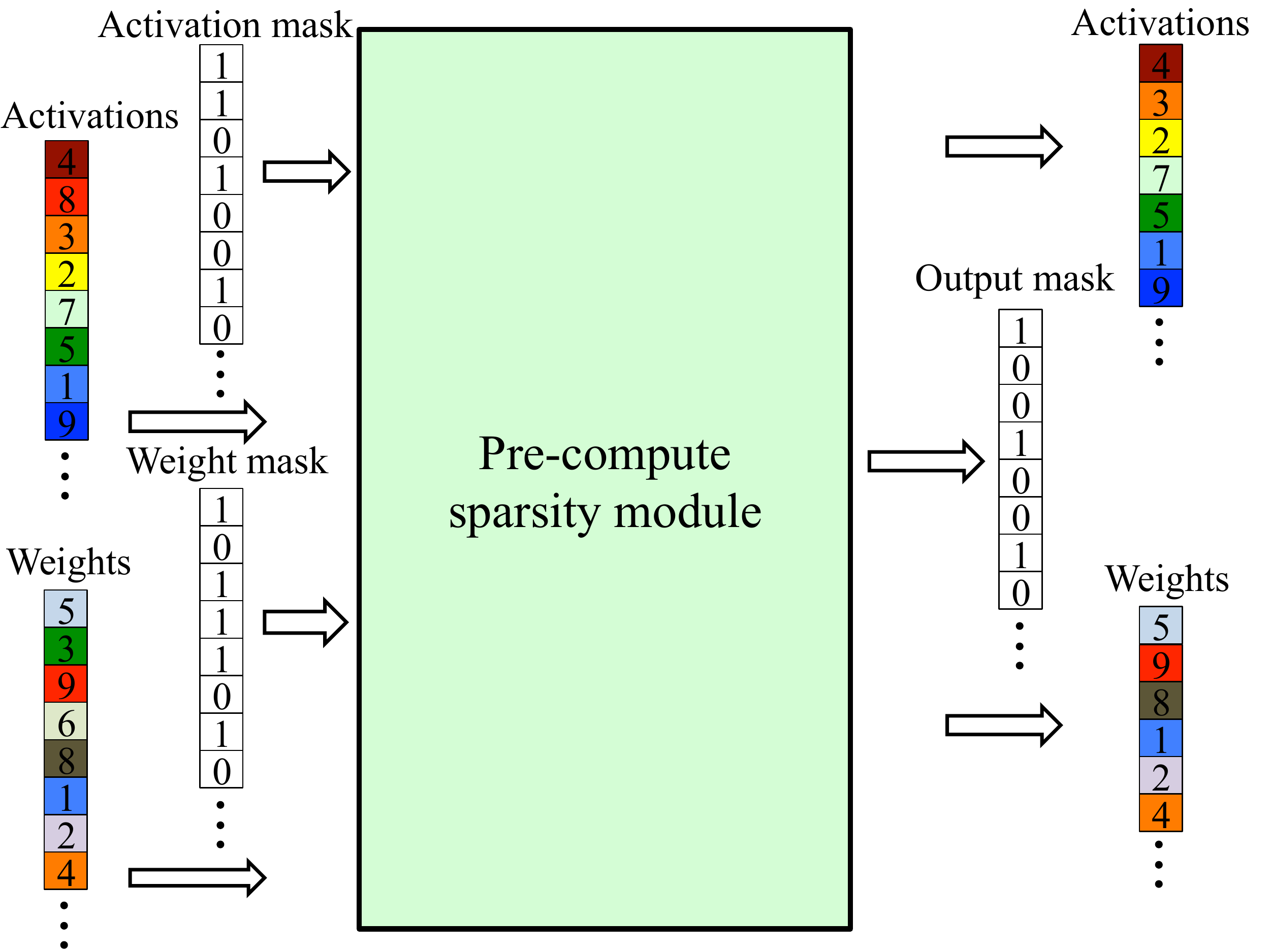}
\caption{The pre-compute sparsity module}
\label{sparse}
\end{figure}

We implement the binary mask scheme using a low overhead pre-compute sparsity module that 
preprocesses the sparse-encoded activations and weights and provides zero-free data to the MAC lanes. 
After output data traverse the MAC lanes, another post-compute sparsity module is used to remove 
all the zero-elements generated by the activation function before storing them back to on-chip 
memory. Fig.~\ref{sparse} shows the pre-compute sparsity module that takes the zero-free data vectors 
and binary mask vectors as inputs, and generates an output mask as well as zero-free
activations/weights for the MAC lanes. The output binary mask indicates the common indexes of 
non-zero elements in both the activation and weight vectors. After being preprocessed by the 
pre-compute sparsity module, the ``dangling" non-zero 
elements in the activation and weight data vectors are removed. The dangling non-zero activations refer to the non-zero elements in the activation data vector where their corresponding weights at the same index are zeros, and vice versa.

\begin{figure*}[!t]
\centering
\includegraphics[width=7.35in]{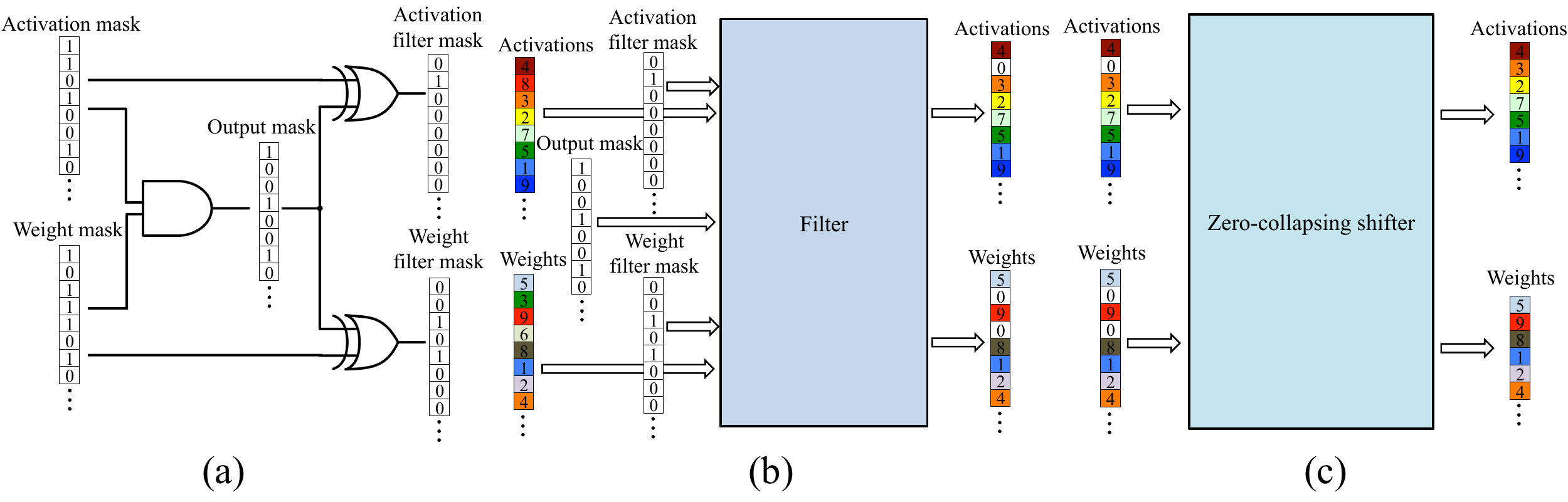}
\caption{The submodules of the pre-compute sparsity module: (a) mask generation, (b) dangling-data filter, and (c) zero-collapsing shifter}
\label{pre}
\end{figure*}

Fig.~\ref{pre}(a) shows the mask generation process used by the pre-compute sparsity module. The 
output mask is the \textbf{AND} of the activation and weight masks. The output mask, together with 
the activation and weight masks, is used by two more \textbf{XOR} gates for filter mask generation. 
Fig.~\ref{pre}(b) shows the dangling data filtering process using the three masks obtained in 
the previous step. The sequential scanning and filtering mechanism for one type of data used in the 
filtering step is shown in Algorithm~\ref{scan}. The data vector, as well as the two mask vectors, 
is scanned in sequence. At each step, a 1 in the output mask implies a common non-zero index. Hence, 
the corresponding element in the data vector passes through the filter. On the other hand, if a 0 
appears in the mask filter and the corresponding mask bit in the filter mask is 1, then a 
dangling non-zero element is detected in the data vector and is blocked by the filter. If both the 
output mask bit and filter mask bit are zeros, it means that the data elements at this index in both 
the activation and weight vectors are zeros and thus already skipped. After filtering out the 
dangling elements in activations and weights, a zero-collapsing shifter is used to remove the 
zeros and keep the data vectors zero-free in a similar sequential scanning manner, as shown in 
Fig.~\ref{pre}(c). These zero-free activations and weights are then fed to the MAC lanes for 
computation. Since only zero-free data are used in the MAC lanes, ineffectual computations are 
completely skipped, thus improving throughput and saving energy.

\begin{algorithm}[t]
\caption{Sequential scanning and filtering mechanism}\label{scan}
\begin{algorithmic}[1]
\State \textbf{Inputs}: in\_data, output\_mask, filter\_mask
\State \textbf{output}: out\_data
\State data\_pointer$\gets$0, mask\_pointer$\gets$0
\While{mask\_pointer $<$ mask\_length}
    \If{output\_mask[mask\_pointer] == 1}
        \State out\_data[data\_pointer] = in\_data[data\_pointer]
        \State data\_pointer++
    \ElsIf{filter\_mask[mask\_pointer] == 1}
        \State out\_data[data\_pointer] = 0
        \State data\_pointer++
    \EndIf
    \State mask\_pointer++
\EndWhile
\end{algorithmic}
\end{algorithm}

\subsection{Reduced-precision processing using stochastic rounding}
SPRING processes CNNs using fixed-point numbers with reduced precision. Every time a new result 
is generated by the CNN, it has to be first rounded to the nearest discrete number, either in 
a floating-point representation or a fixed-point representation. Since the gap between adjacent 
numbers in the fixed-point representation is much larger than in the floating-point representation, 
the resulting quantization error in the former is much more pronounced. This prevents the fixed-point 
representation from being used in error-sensitive CNN training. In order to utilize the faster and 
more energy-efficient fixed-point arithmetic units, we adopt the \textit{stochastic rounding} method 
proposed in \cite{sr}. The traditional deterministic rounding scheme always rounds a real number to 
its nearest discrete number, as shown in Eq.~\ref{round}. We follow the 
definitions used in \cite{sr}, where $\epsilon$ denotes the
smallest positive discrete number supported in the fixed-point format
and $\floor{x}$ is defined as the largest integer multiple of $\epsilon$ less 
than or equal to x.

\begin{equation}\label{round}
    Round(x) =
    \begin{cases}
        \floor{x} & {if \ x < \floor{x} + \dfrac{\epsilon}{2}}\\
        \floor{x} + \epsilon & {otherwise}
    \end{cases}
\end{equation}

In contrast, a real number $x$ is rounded to $\floor{x}$ and $\floor{x} + \epsilon$ stochastically in the 
stochastic rounding scheme, as shown in Eq.~\ref{rounding} \cite{sr}. It is shown in \cite{sr} that 
with the stochastic rounding scheme, the CNN weights can be trained to tolerate the quantization 
noise without increasing the number of cycles required for convergence.

\begin{equation}\label{rounding}
    Round(x) = 
    \begin{cases}
    \floor{x} & {with \ probability \  \dfrac{\floor{x} + \epsilon - x}{\epsilon}}\\
    \floor{x} + \epsilon & {with \ probability \ \dfrac{x-\floor{x}}{\epsilon}}
    \end{cases}
\end{equation}

\begin{figure}[!t]
\centering
\includegraphics[width=3.5in]{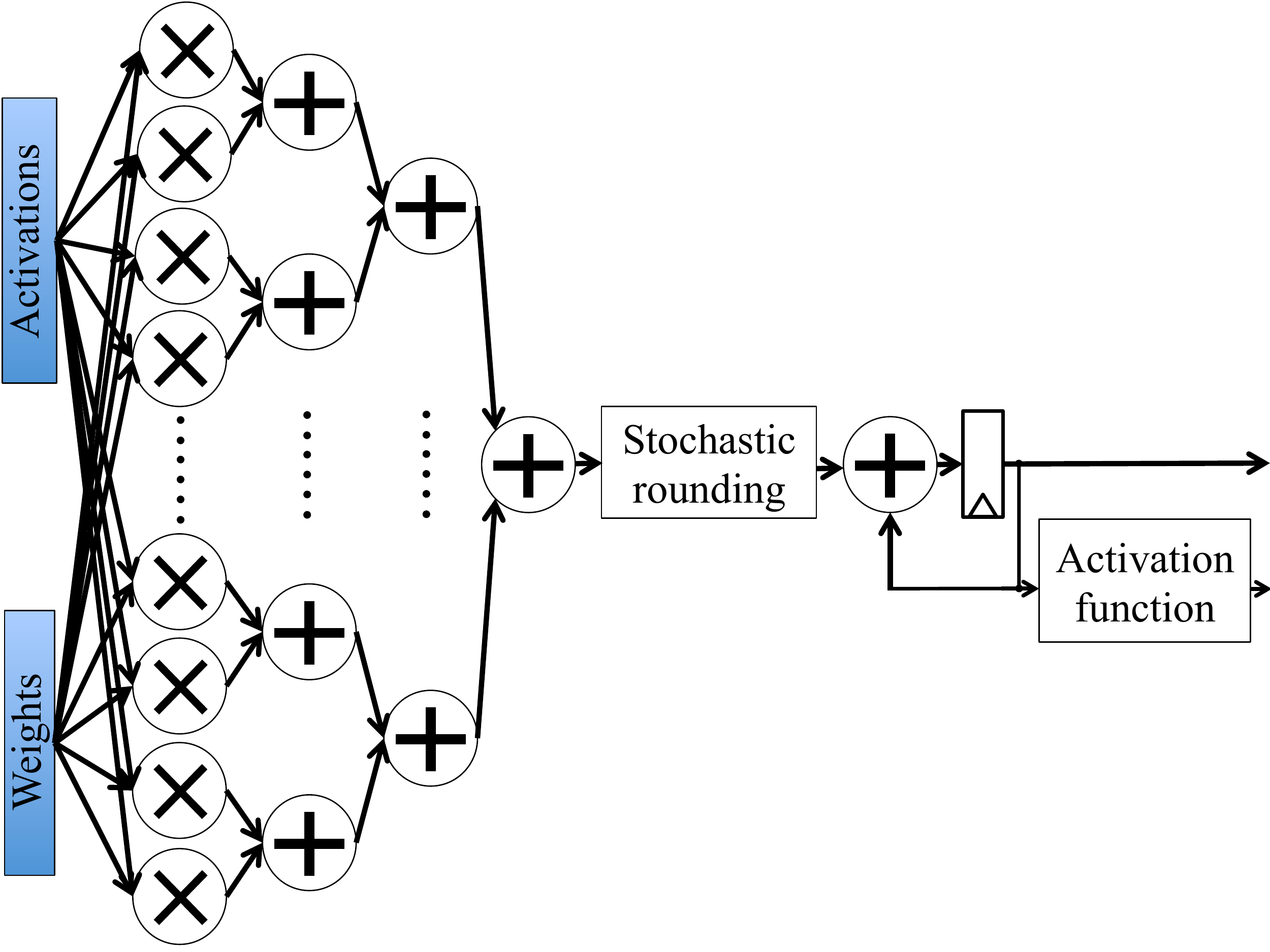}
\caption{The MAC lane}
\label{mac}
\end{figure}

The stochastic rounding scheme is embedded in the MAC lane, as shown in Fig.~\ref{mac}. Activations 
and weights are represented using fixed-point numbers using IL+FL bits, where IL denotes the number 
of bits for the integer portion and FL denotes the number of bits for the fraction part. The 
zero-free activations and weights from the pre-compute sparsity module are subject to
multiplications in the MAC lanes, where the products are represented with 2$\times$IL integer bits 
and 2$\times$FL fractional bits to prevent overflow. Accumulations over products are also performed 
using 2$\times$(IL+FL) bits. Then, a stochastic rounding module is used to reduce the numerical 
precision before applying the activation function or storing the result back to on-chip memory. We 
use a linear-feedback shift register to generate pseudo-random numbers for stochastic rounding.

\begin{figure*}[!t]
\centering
\includegraphics[width=7.2in]{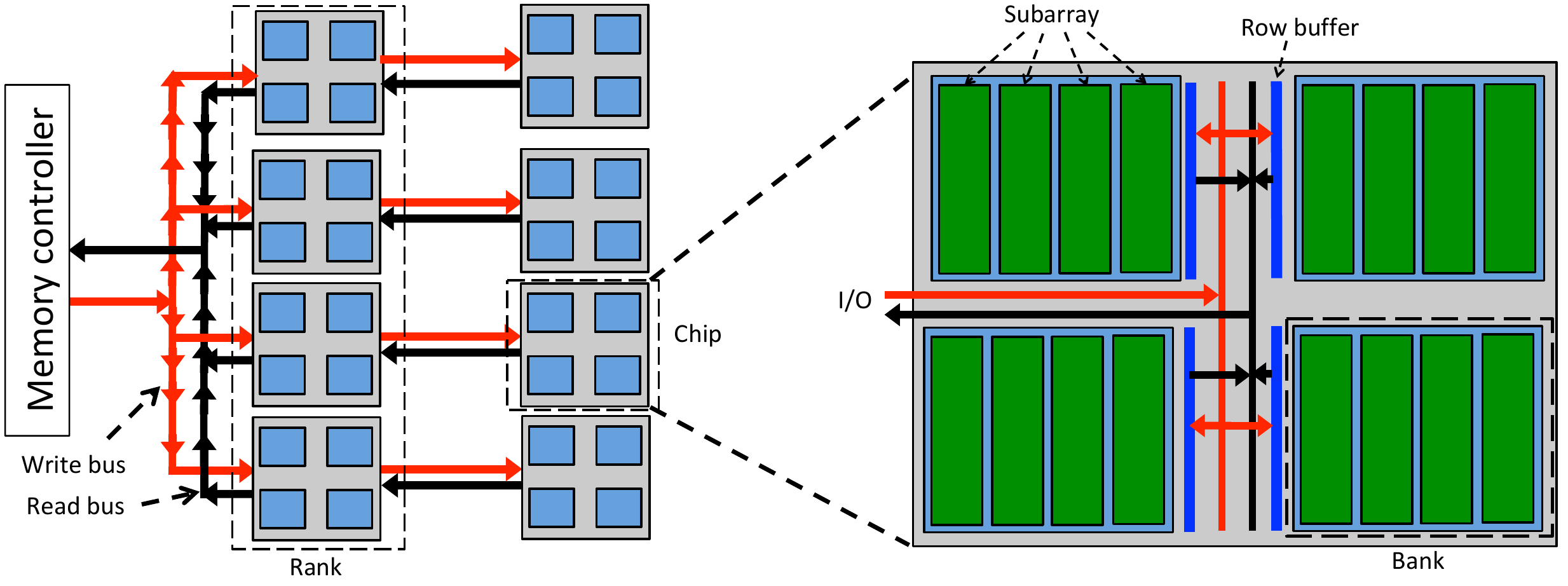}
\caption{Read/write decoupled interconnects \cite{hybrid}}
\label{interconnect}
\end{figure*}

\subsection{Monolithic 3D NVRAM interface}
SPRING uses a monolithic 3D NVRAM interface previously proposed in \cite{monolithic} and adapts
it to its 3D architecture to provide the accelerator tier with significant memory bandwidth.
As shown in Fig.~\ref{spring}, SPRING uses two memory channels where each channel has its own
memory controller to control the associated two RRAM ranks. An ultra-wide memory bus (1KB wide)
is used in each channel, since the interconnects between SPRING and memory controllers, and between 
memory controllers and RRAM ranks, are implemented using vertical MIVs. This on-chip memory bus not 
only reduces the access latency relative to the conventional off-chip memory bus, but also makes 
row-wide granular memory accesses possible to enable energy savings. In addition, the column decoder 
can be removed to reduce the access latency and power dissipation in this row-wide access 
granularity scheme. To reduce repeated accesses to the same row, especially the energy-consuming 
write accesses of RRAM, the row buffer is reused as the write buffer. A dirty bit is used to indicate 
if the corresponding row entry in the row buffer needs to be written back to the RRAM array when 
flushed out. The read and write accesses are decoupled by adding another set of vertical 
interconnects, as shown in Fig.~\ref{interconnect} \cite{hybrid}. Hence, the slower write access 
does not block the faster read access and thus a higher memory bandwidth is achieved. In addition, 
RRAM nonvolatility not only enables the elimination of bulky periodic refresh circuitry, but also 
allows the RRAM arrays to be powered down in the idle intervals to reduce leakage power. A rank-level 
adaptive power-down policy is used to maintain a balance between performance and energy saving: the 
power-down threshold for each RRAM rank is adapted to its idling pattern so that a rank is only 
powered down if it is expected to be idle for a long time.

\section{Simulation methodology}
\label{simulations}
In this section, we present the simulation flow for SPRING and the experimental setup.

\begin{figure}[!t]
\centering
\includegraphics[width=3.5in]{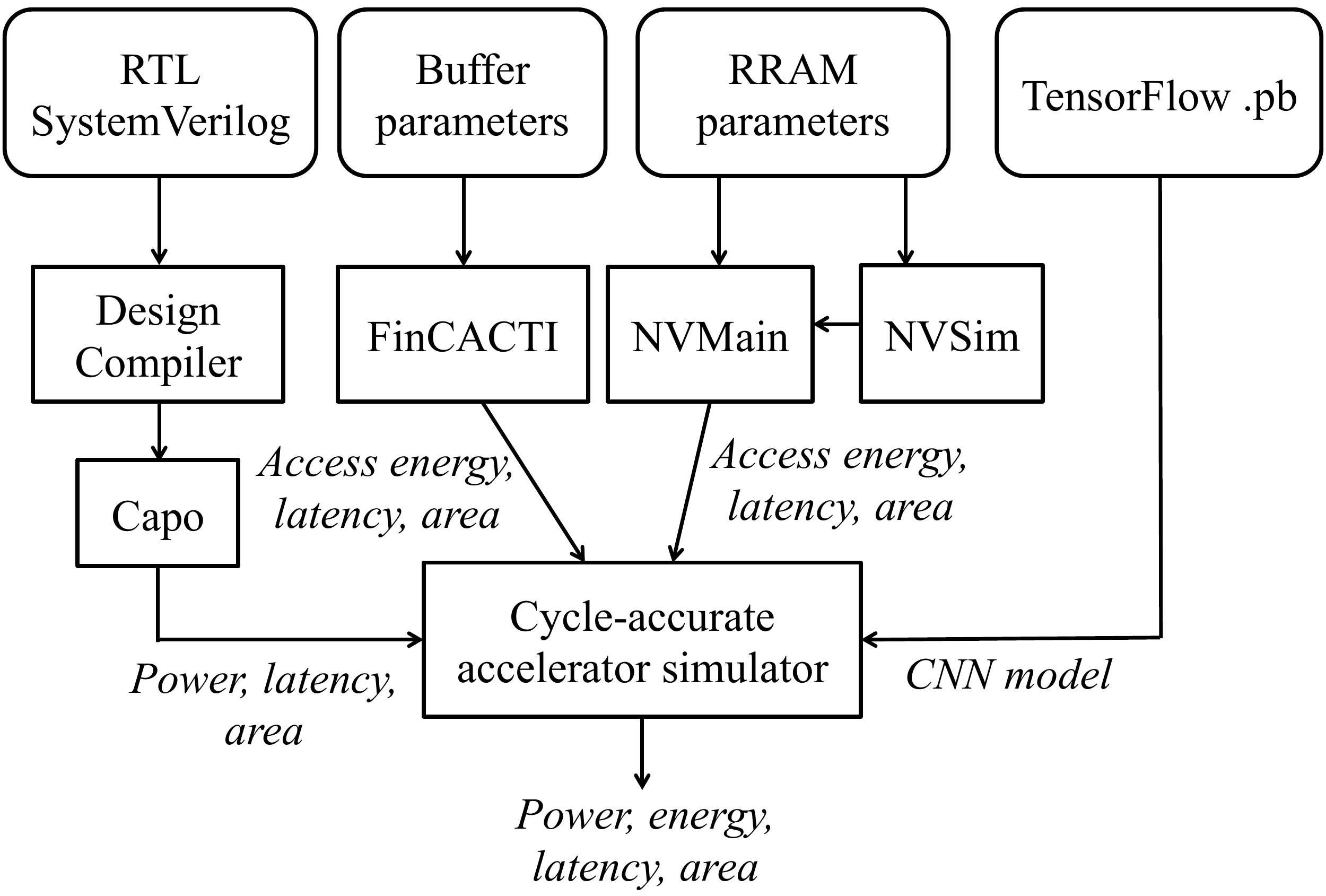}
\caption{Simulation flow}
\label{flow}
\end{figure}

Fig.~\ref{flow} shows the simulation flow used to evaluate the proposed SPRING accelerator
architecture. We implement components of SPRING at the register-transfer level (RTL) with 
SystemVerilog to estimate delay, power, and area. The RTL design is synthesized by Design Compiler 
\cite{compiler} using a 14nm FinFET technology library \cite{14nm}. Floorplanning is done by Capo 
\cite{capo}, an open-source floorplacer. On-chip buffers are modeled using FinCACTI \cite{fincacti}, a
cache modeling tool enhanced from CACTI \cite{cacti}, to support deeply-scaled FinFETs at the 14nm
technology node. The monolithic 3D RRAM system is modeled by NVSim \cite{NVSim}, a
circuit-level memory simulator for emerging NVRAMs, and NVMain \cite{NVMain}, an emerging NVRAM
architecture simulator. The synthesized results, together with buffer and RRAM estimations, are
then plugged into a customized cycle-accurate Python simulator. This accelerator simulator
takes CNNs in the TensorFlow \cite{tensorflow} Protocol Buffers format and estimates the
computation latency, power dissipation, energy consumption, and area. SPRING treats the TensorFlow 
operations like complex instruction set computer instructions where each operation involves 
many low-level operations and the CNNs are mapped to SPRING using an analytical model similar to the one used in \cite{ali}.

We compare our design with the Nvidia GeForce GTX 1080 Ti GPU, which uses the Pascal microarchitecture 
\cite{pascal} in a 16nm technology node. The die size of GTX 1080 Ti is 471$mm^2$ and the base 
operating frequency is 1.48 $GHz$, which can be boosted to 1.58 $GHz$. GTX 1080 Ti uses an 11 GB 
GDDR5X memory with 484 GB/s memory bandwidth to provide 10.16 TFLOPS peak single-precision 
performance.

We evaluate SPRING and GTX 1080 Ti on seven well-known CNNs: Inception-Resnet V2 
\cite{inceptionresnet}, Inception V3 \cite{inception}, MobileNet V2 \cite{mobilenetv2}, 
NASNet-mobile \cite{nasnet}, PNASNet-mobile \cite{pnasnet}, Resnet-152 V2 \cite{resnet}, and 
VGG-19 \cite{vgg}. We evaluate both the training and inference phases of these CNNs on the ImageNet 
dataset \cite{imagenet}. The sparsity of the CNNs are assumed to be 50\%, as it is shown in \cite{dma} that the average sparsity level of widely used CNNs, such as AlexNet \cite{alexnet}, VGG \cite{vgg}, and Inception \cite{inception}, are over 50\%.  We use the default batch sizes defined in the TensorFlow-Slim library 
\cite{slim}: 32 for training and 100 for inference.

\section{Experimental results}
\label{results}
In this section, we present experimental results for SPRING and compare them with those for GTX 
1080 Ti.

Table~\ref{parameters} shows the values of various design parameters used in SPRING. They are 
obtained through the accelerator design space exploration methodology proposed in \cite{ali}. It is shown 
in \cite{sr} that with 16 FL bits, training CNNs using the stochastic rounding scheme can converge 
in a similar amount of time with a negligible accuracy loss relative to when single-precision 
floating-point arithmetic is used. Hence, we use 4 IL bits and 16 FL bits in the fixed-point 
representation. The convolution loop order refers to the execution order of the multiple for-loops in 
the CONV layer. SPRING executes convolutions by first unrolling the for-loops across multiple inputs 
in the batch. Then, it unrolls the for-loops within the filter weights, followed by unrolling in the 
activation channel dimension. In the next step, it unrolls the for-loops with activation feature 
maps. Finally, it unrolls for-loops across the output channels.

\begin{figure}[!t]
\centering
\includegraphics[width=3.5in]{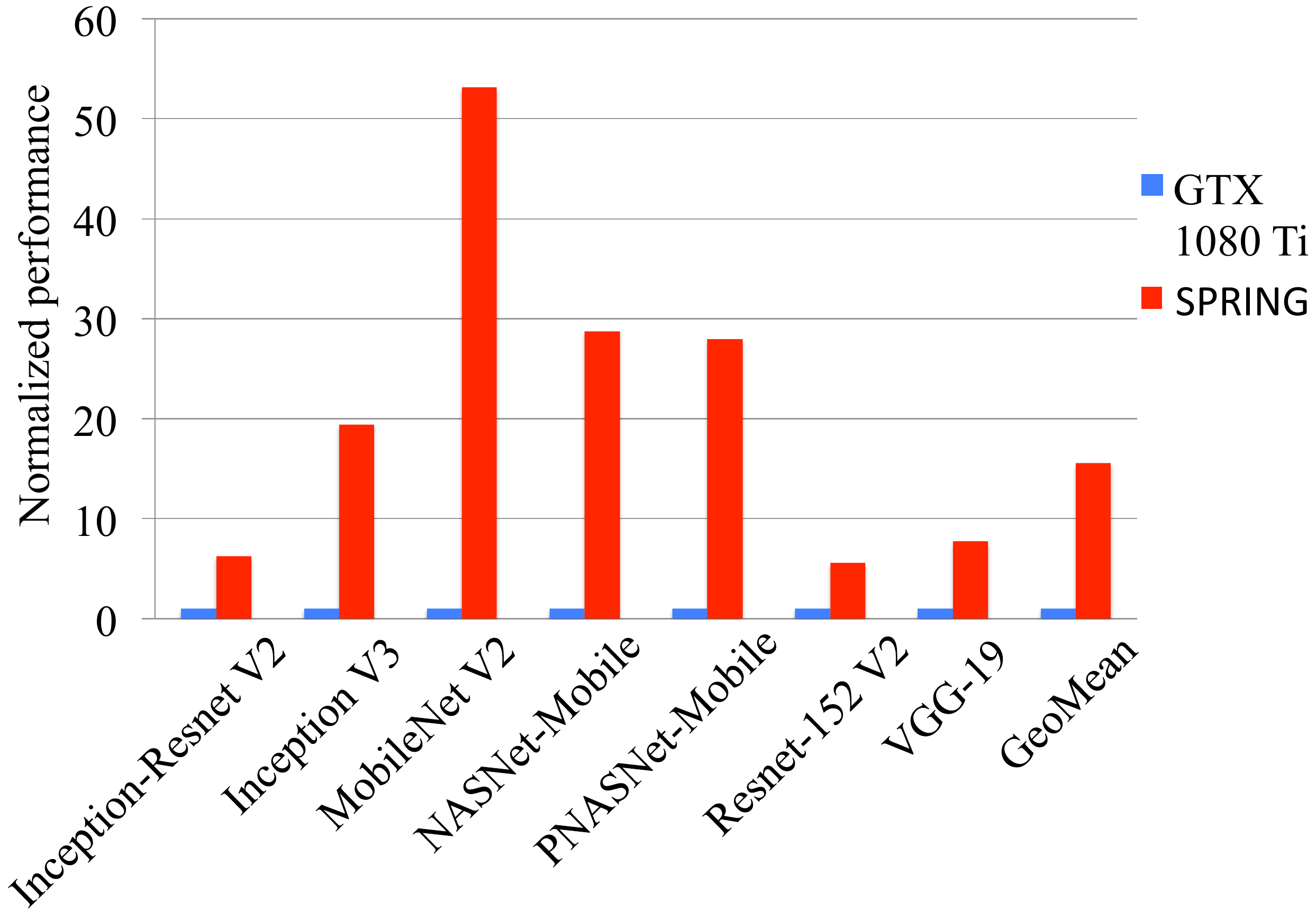}
\caption{Normalized training performance}
\label{training_performance}
\end{figure}

\begin{figure}[!t]
\centering
\includegraphics[width=3.5in]{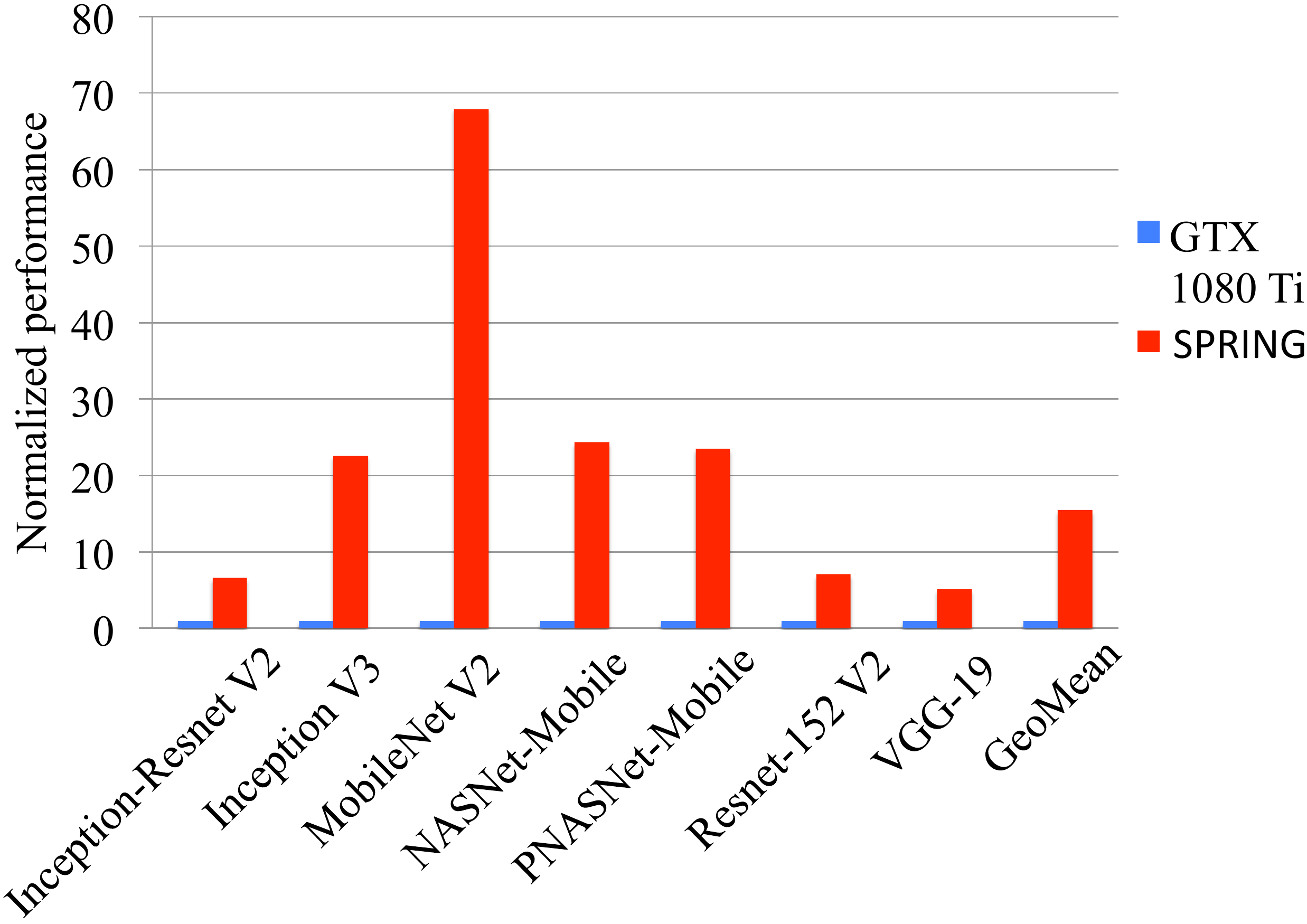}
\caption{Normalized inference performance}
\label{inference_performance}
\end{figure}

\begin{table}[t]
\centering
\caption{SPRING design parameters}
\label{parameters}
\scalebox{0.9}{
\begin{tabular}{*{2}{|c}|}
\hline
Accelerator parameters & Values\\
\hline \hline
Clock rate & 700 $MHz$\\
\hline
Number of PEs & 64\\
\hline
Number of MAC lanes per PE & 72\\
\hline
Number of multipliers per MAC lane & 16\\
\hline
Weight buffer size & 24 MB\\
\hline
Activation buffer size & 12 MB\\
\hline
Mask buffer size & 4 MB\\
\hline
Convolution loop order & batch-weight-in channel-input-out channel\\ 
\hline
IL bits & 4\\
\hline
FL bits & 16\\
\hline
Technology & 14nm FinFET\\
\hline
Area & 151 $mm^2$\\
\hline
Monolithic 3D RRAM & 8 GB, 2 channels, 2 ranks, 16 banks,\\
& 1 KB bus, $t_{BURST}$=0.5 ns, 2.0 $GHz$ \cite{5ns}\\ 
\hline
\end{tabular}
}
\end{table}

Fig.~\ref{training_performance} and Fig.~\ref{inference_performance} show the normalized
performance of SPRING and GTX 1080 Ti over the seven CNNs in the training and inference
phases, respectively. All results are normalized to those of GTX 1080 Ti. In the training phase, SPRING achieves speedups 
ranging from 5.5$\times$ to 53.1$\times$ with a geometric mean of 15.6$\times$ on the seven
CNNs. In the inference phase, SPRING is faster than GTX 1080 Ti by 5.1$\times$ to
67.9$\times$ with a geometric mean of 15.5$\times$. In both cases, SPRING has better
performance speedups on relatively light-weight CNNs, i.e., MobileNet V2, NASNet-mobile, and
PNASNet-mobile. This is because these light-weight CNNs do not require large volumes of
activations and weights to be transferred between the external memory and on-chip buffers. Therefore,
the memory bandwidth bottleneck is alleviated and the speedup from sparsity-aware
computation becomes more noteworthy. On the other hand, on large CNNs, such as Inception-Resnet
V2 and VGG-19, the sparsity-aware MAC lanes of SPRING idle and wait for data fetch from the RRAM 
system, lowering the performance speedup relative to GTX 1080 Ti.

\begin{figure}[!t]
\centering
\includegraphics[width=3.5in]{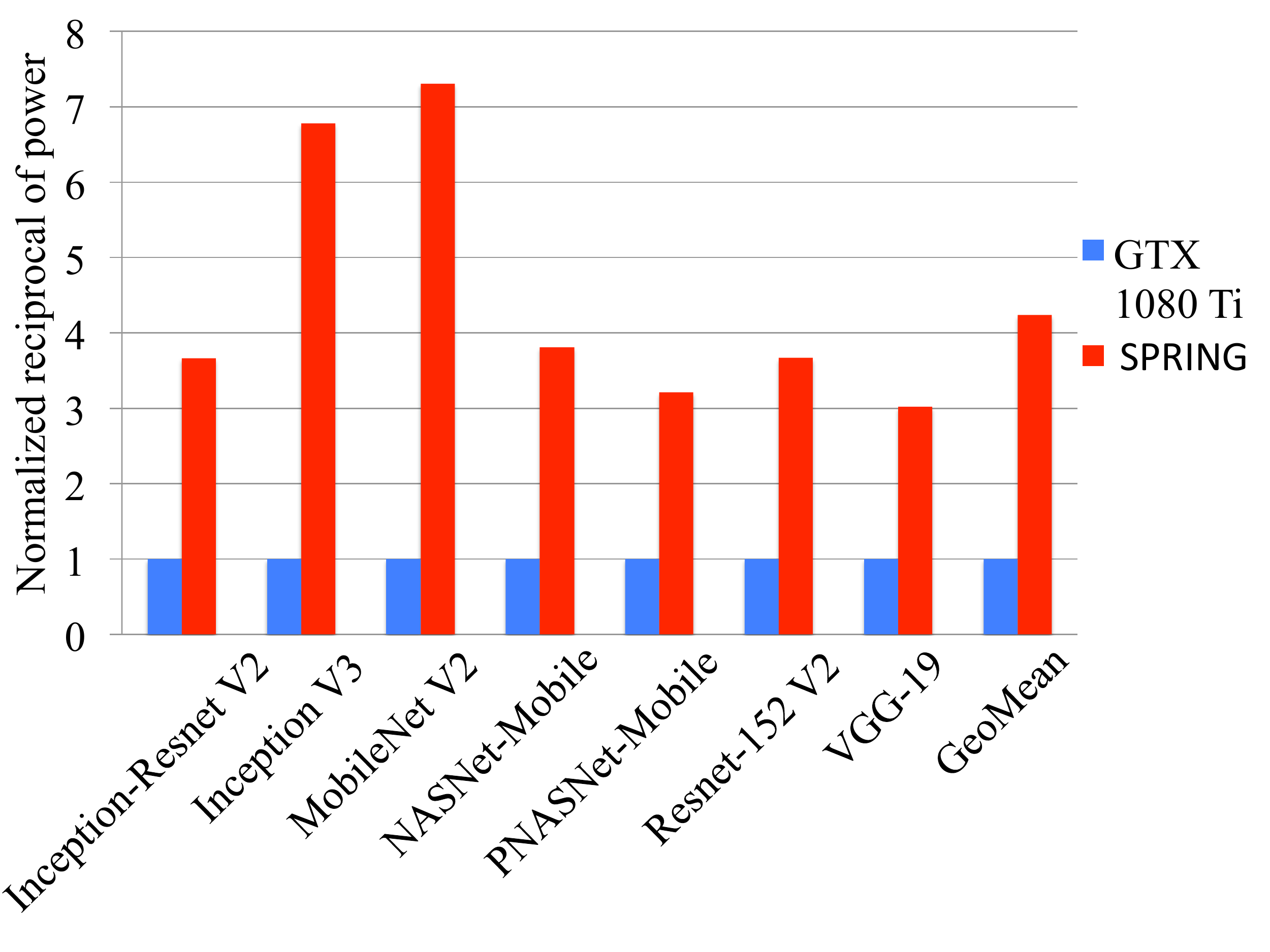}
\caption{Normalized reciprocal of power for training}
\label{training_power}
\end{figure}

\begin{figure}[!t]
\centering
\includegraphics[width=3.5in]{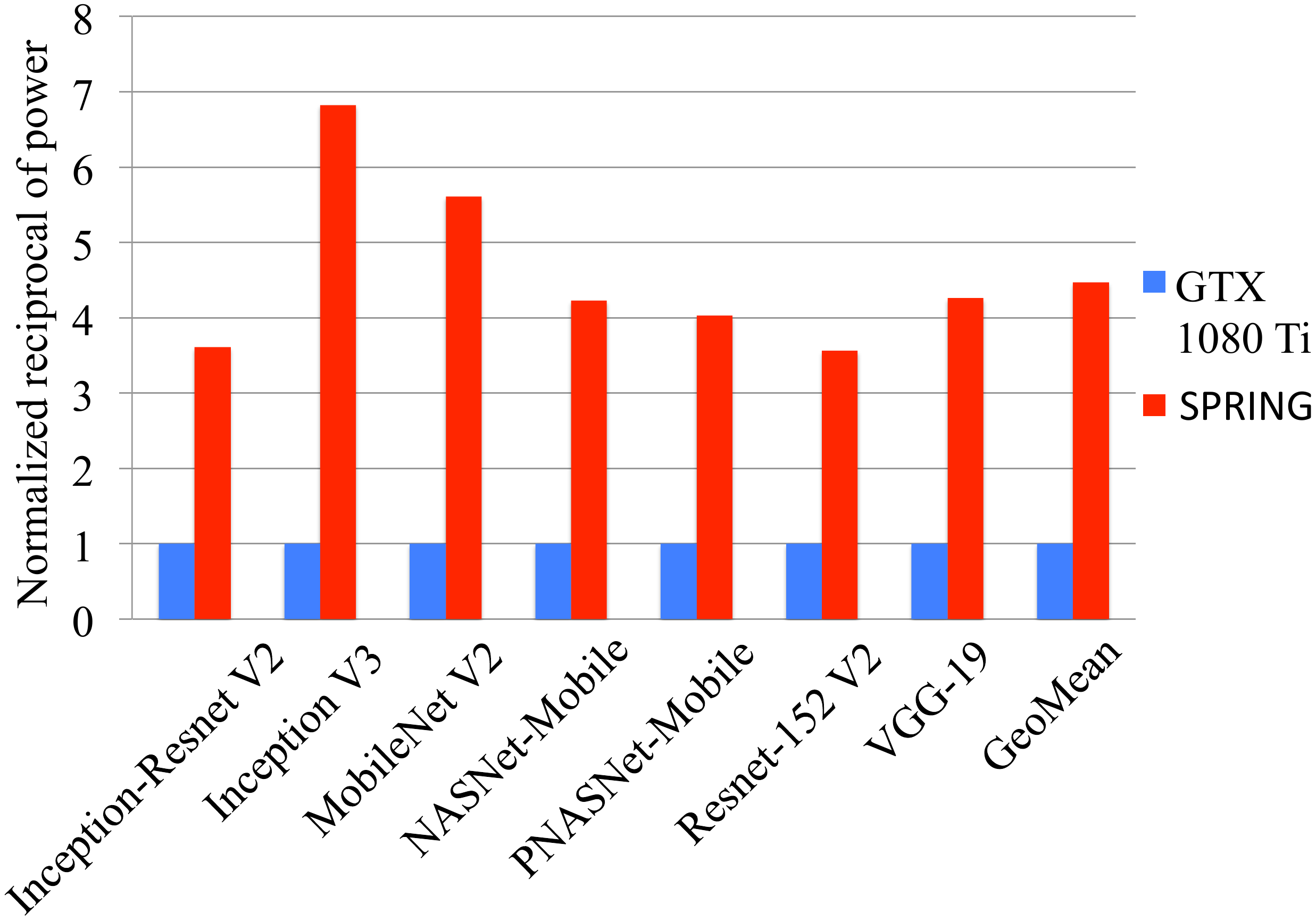}
\caption{Normalized reciprocal of power for inference}
\label{inference_power}
\end{figure}

Fig.~\ref{training_power} and Fig.~\ref{inference_power} show the normalized reciprocal of power of 
SPRING and GTX 1080 Ti in training and inference, respectively. All results are normalized to those of GTX 1080 Ti. On an average, SPRING reduces 
power dissipation by 4.2$\times$ and 4.5$\times$ for training and inference, respectively.

\begin{figure}[!t]
\centering
\includegraphics[width=3.5in]{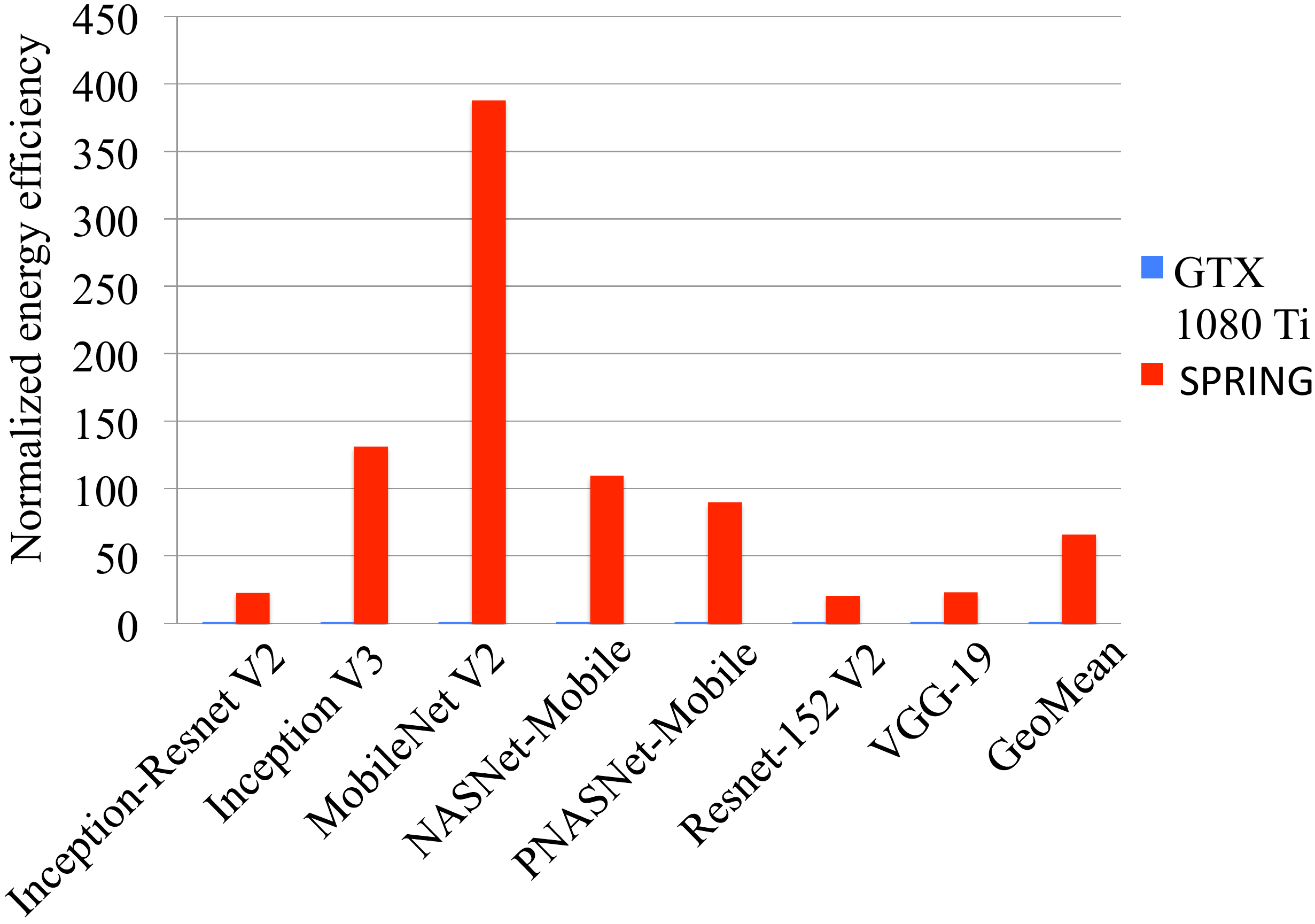}
\caption{Normalized energy efficiency in training}
\label{training_energy}
\end{figure}

\begin{figure}[!t]
\centering
\includegraphics[width=3.5in]{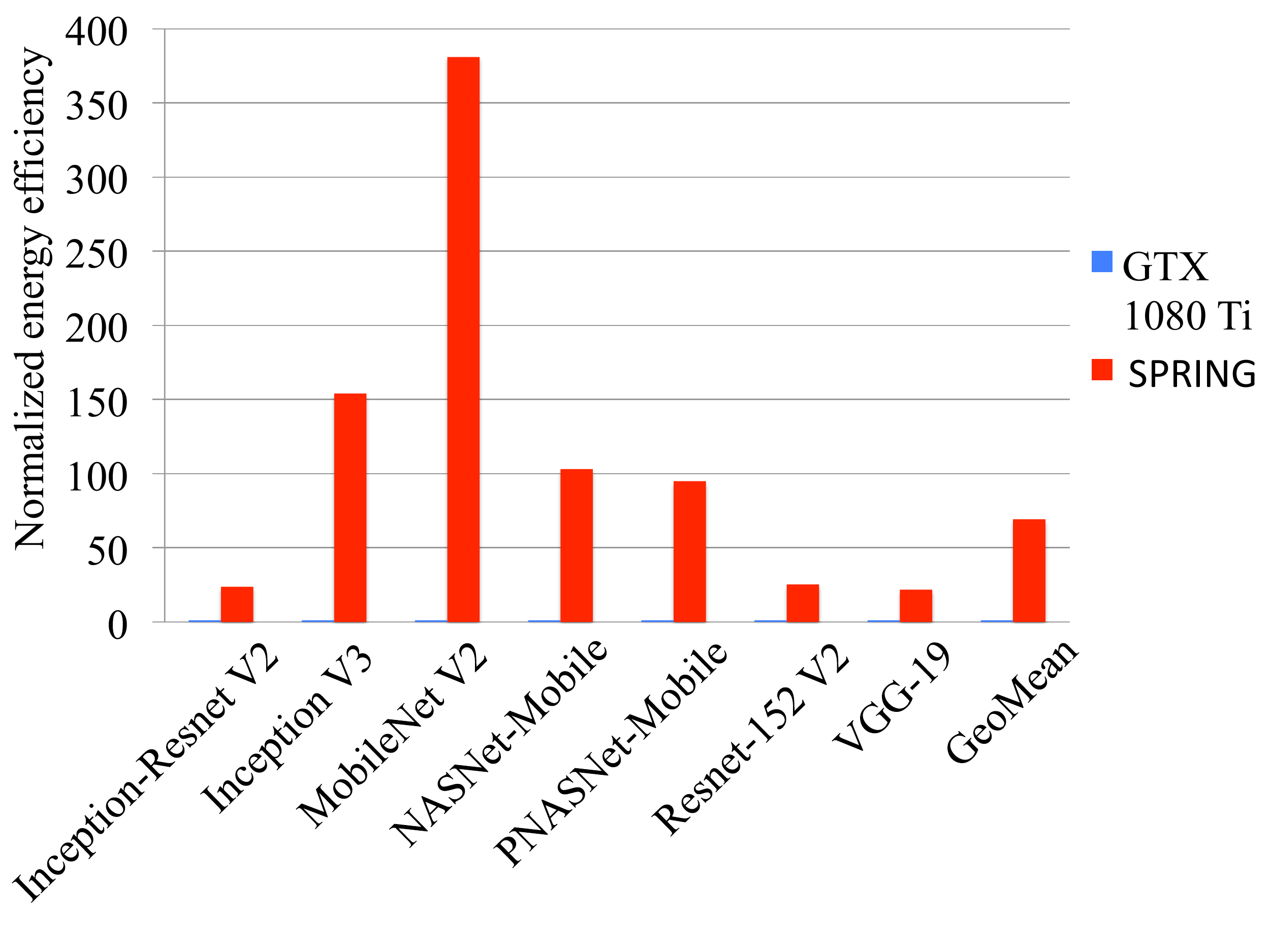}
\caption{Normalized energy efficiency in inference}
\label{inference_energy}
\end{figure}

Fig.~\ref{training_energy} and Fig.~\ref{inference_energy} show the normalized energy efficiency of 
SPRING and GTX 1080 Ti for training and inference, respectively. All results are normalized to those of GTX 1080 Ti. Compared to the GTX 1080 Ti, 
SPRING achieves an average of 66.0$\times$ and 69.1$\times$ energy efficiency improvements in
training and inference, respectively. This makes the GTX 1080 Ti columns invisible. We observe that, among the seven CNNs, SPRING achieves
the best normalized energy efficiency on MobileNet V2, both in the training and inference
phases. Since MobileNet V2 has a much smaller network size (97.6\% parameter reduction compared
to VGG-19 \cite{mobilenetv2}), most of the network weights can be retained in on-chip buffers
without accessing the external memory. Hence, SPRING can reduce energy consumption
significantly through our sparsity-aware acceleration scheme. On the other hand, energy reduction 
from sparsity-aware computation is offset by energy-consuming memory accesses on large CNNs, such 
as Inception-Resnet V2 and VGG-19. This is consistent with the results reported in \cite{cambriconx} 
that show that over 80\% of the total energy consumption is from memory access.

\section{Discussions and limitations}
\label{discussion}
In this section, we discuss the assumptions we made in this work and the limitations of the
SPRING architecture.

The SPRING architecture is based on the assumption that the monolithic
3D stacking technology is emerging in the near future. One main
disadvantage of monolithic 3D integration, compared to the TSV-based 3D
stacking, is that the high-temperature process used for the top tiers
may damage the interconnects on the tiers below. To prevent or minimize
this damage, either low-temperature annealing is used for the top tiers
or Copper is replaced with Tungsten as the interconnect material on the 
bottom tiers. However, low-temperature annealing may degrade device 
performance on the top tiers while the Tungsten interconnect is less 
competitive with the Copper counterpart in terms of performance on the bottom 
tiers. However, it is still not clear how the low thermal budget process on 
the top tiers correlates to the degree of top-tier device degradation for 
advanced technology nodes \cite{Ku:2016}. Some previous works assume the 
devices on the top tiers will be degraded to some extent and evaluate the 
overall chip performance based on this assumption 
\cite{Samal:2016,Panth:2014}. On the other hand, there are works suggesting 
that the degradation is negligible when certain monolithic 3D processes 
are used \cite{liu2012ultra,CEA}. For example, it is shown in \cite{LETI} 
that using Tungsten for interconnects on the bottom tiers for 
interconnect-dominant circuits has a limited impact on overall performance 
and power consumption (less than 2\% degradation in both performance and power 
consumption). Hence, we do not model the degradation from the monolithic 3D 
integration process in this work.

The performance speedup, power reduction ratio, and energy efficiency improvement reported in 
Section~\ref{results} are obtained at the batch level. We use batch-level training results 
since the CNN training results are based on the assumption that with sufficient precision bits, 
fixed-point training using stochastic rounding scheme can lead to convergence with no worse number 
of cycles than the training process based on single-precision floating-point arithmetic, as 
suggested in \cite{sr}, where 16 FL bits are used for fixed-point training with stochastic rounding 
and the convergent epoch number is similar to that of single-precision floating-point training.

A major limitation of the SPRING accelerator architecture is that the sequential scanning and
filtering mechanism shown in Algorithm~\ref{scan} needs multiple cycles to filter out dangling
non-zero elements and collapse the resulting zeros. This may incur a long latency in data
preprocessing, which makes SPRING unsuitable for latency-sensitive edge inference applications. 
However, since this sequential scanning and filtering scheme is pipelined, the overall throughput is 
unaffected and therefore the total latency for one batch is independent of the sequential scanning 
steps used by the pre-compute sparsity module.

Our binary mask encoding method is similar to the \textit{dual indexing} encoding proposed in 
\cite{DIM}. Although we both use a binary mask to point to the index of non-zero elements in the 
data vector, our binary mask encoding scheme has several advantages. First, the index masks are 
kept in binary form throughout the entire sparsity encoding and decoding process. Hence, the storage 
overhead of the binary mask is at most 5\%, assuming 4 IL bits and 16 FL bits. The real storage 
overhead is much lower than this value since most of activations and weights are zeros. However, 
the binary masks are converted to decimal masks in \cite{DIM} to serve as select signals of
a MUX. This not only increases the storage overhead of the masks, but also increases the computation 
complexity of mask manipulation. Besides, their binary-to-decimal mask transfer process is 
sequential, which incurs a long processing latency that increases as the size of the mask vector 
increases.

While preparing this article, we became aware of a very recent work that 
shares our motivations but adopts different approaches. Eager 
Pruning \cite{EP}, an algorithm-architecture co-design method, speeds up 
DNN training by moving pruning to the training phase. It is observed 
in \cite{EP} that the ranking of weight magnitudes is relatively stable during 
the training process. Hence, insignificant weights can be identified and 
pruned in the early training stage. This reduced training computation is 
then transformed into speedup through a dedicated accelerator architecture. 
The sparse weights are distributed to multiple PEs using a Dynamically 
Reconfigurable Add and Collect Tree (DRACT) to support the dataflow of 
Eager Pruning. However, Eager Pruning only supports speedup from weight 
sparsity but lacks support for activation sparsity. The sparsity-aware 
acceleration scheme discussed in this article can be combined with the Eager 
Pruning dataflow through the help of DRACT for the full use of data sparsity 
(both activations and weights).

\section{Conclusion}
\label{conclusion}
In this article, we proposed a sparsity-aware reduced-precision CNN accelerator, named SPRING. A
binary mask scheme is used to encode weight/activation sparsity.  It is efficiently processed 
through a sequential scanning and filtering mechanism. SPRING adopts the stochastic rounding 
algorithm to train CNNs using reduced-precision fixed-point numerical representation. An efficient
monolithic 3D NVRAM interface is used to provide significant memory bandwidth for CNN
evaluation. Compared to Nvidia GeForce GTX 1080 Ti, SPRING achieves 15.6$\times$, 4.2$\times$, 
and 66.0$\times$ improvements in performance, power reduction, and energy efficiency, respectively, 
in the training phase, and 15.5$\times$, 4.5$\times$, and 69.1$\times$ improvements in performance, 
power reduction, and energy efficiency, respectively, in the inference phase.

\ifCLASSOPTIONcaptionsoff
  \newpage
\fi

\bibliographystyle{IEEEtran}

\bibliography{bare_adv}

\begin{IEEEbiography}[{\includegraphics[width=1in,height=1.25in,clip,keepaspectratio]{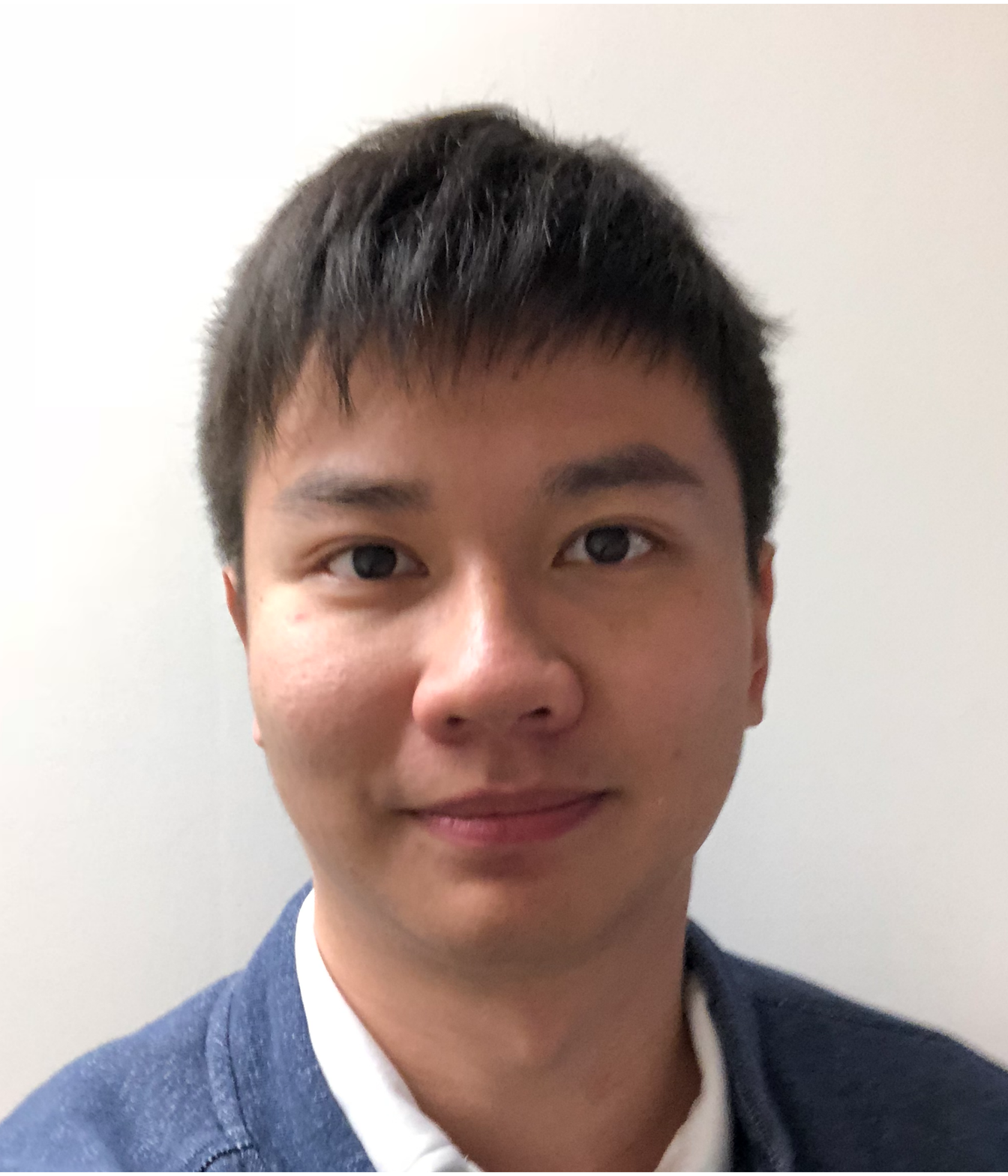}}]
{Ye Yu} received the B.Eng. degree in Electronic and Computer Engineering from 
The Hong Kong University of Science and Technology, Hong Kong, China, in 2014, 
and the M.A. and Ph.D. degrees in Electrical Engineering from Princeton 
University, NJ, USA, in 2016 and 2019, respectively. He is currently a 
software engineer at Microsoft.

His current research interests include computer vision, machine learning, 
and deep learning model compression and acceleration.
\end{IEEEbiography}

\begin{IEEEbiography}[{\includegraphics[width=1in,height=1.25in,clip,keepaspectratio]{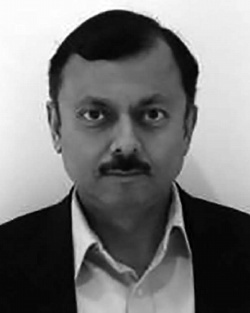}}]
{Niraj K. Jha} (S'85-M'85-SM'93-F'98) 
received his B.Tech. degree in Electronics and Electrical
Communication Engineering from Indian Institute of Technology,
Kharagpur, in 1981 and Ph.D. degree in Electrical Engineering from
University of Illinois at Urbana-Champaign, IL in 1985. He has been a
faculty member of the Department of Electrical Engineering, Princeton
University, since 1987. He is a Fellow of IEEE and ACM, and was given
the Distinguished Alumnus Award by I.I.T., Kharagpur. He has also
received the Princeton Graduate Mentoring Award.

He has served as the Editor-in-Chief of IEEE Transactions on VLSI
Systems and as an Associate Editor of several other journals. He has
co-authored five books that are widely used. His research has won 20
best paper awards or nominations.  His research interests include smart
healthcare, cybersecurity, machine learning, and monolithic 3D IC
design. He has given several keynote speeches in the area of
nanoelectronic design/test and smart healthcare. 

\end{IEEEbiography}

\end{document}